\documentclass[%
 aip,
 amsmath,amssymb,
 reprint,%
]{revtex4-2}

\usepackage[utf8]{inputenc}
\usepackage[T1]{fontenc}
\usepackage{mathptmx}
\usepackage{etoolbox}

\usepackage[dvipsnames]{xcolor}
\usepackage{graphicx}
\usepackage{dcolumn}
\usepackage{bm}
\newcommand{\eps}{\epsilon\epsilon_0}
\newcommand{\dd}{\mathrm{d}}
\newcommand{\rb}{\mathbf{r}}

\newcommand{\F}{\mathcal{F}}

\usepackage{ulem}
\usepackage{mhchem}

\definecolor{green}{rgb}{0.1, 0.5, 0.2}

\makeatletter
\def\@email#1#2{%
 \endgroup
 \patchcmd{\titleblock@produce}
  {\frontmatter@RRAPformat}
  {\frontmatter@RRAPformat{\produce@RRAP{*#1\href{mailto:#2}{#2}}}\frontmatter@RRAPformat}
  {}{}
}%
\makeatother
\begin{document}

\preprint{AIP/123-QED}

\title[The Differential Capacitance as a probe for the electric double layer structure and the electrolyte bulk composition]{The Differential Capacitance as a probe for the electric double layer structure and the electrolyte bulk composition}
\author{Peter Cats}
 \email{p.cats@uu.nl}
 \affiliation{Institute for Theoretical Physics, Center for Extreme Matter and Emergent Phenomena, Utrecht University, Princetonplein 5, 3584 CC, the Netherlands}
\author{René van Roij}%
 \email{r.vanroij@uu.nl}
\affiliation{Institute for Theoretical Physics, Center for Extreme Matter and Emergent Phenomena, Utrecht University, Princetonplein 5, 3584 CC, the Netherlands}%

\date{\today}

\begin{abstract}
In this work we theoretically study the differential capacitance of an aqueous electrolyte in contact with a planar electrode, using classical Density Functional Theory, and show how this measurable quantity can be used as a probe to better understand the structure and composition of the electric double layer at play. Specifically, we show how small trace amounts of divalent ions can influence the differential capacitance greatly, and also how small ions dominate its behaviour for  high electrode potentials. In this study, we consider primitive model electrolytes, and not only use the standard definition of the differential capacitance, but also derive a new expression from mechanical equilibrium in a planar geometry. This expression reveals explicitly that the first layer of ions near the charged surface is key to its understanding.  Our insights might be used as a guide in experiments to better understand the electrolyte-electrode interface as well as the (composition of the) bulk electrolyte.
\end{abstract}

\maketitle

\section{Introduction}

At the moment, energy storage is one of the major challenges in the battle against climate change. There are many ``green" methods nowadays to transform one energy form into another~\cite{energy_transition}, with electricity as the favoured end product, but the options to store it efficiently on a large scale are limited. With solar and wind energy on the rise, the era of fossil energy sources will come to and end. However, sunshine and wind are intermittent and therefore the amount of harvested energy is not constant and can fluctuate quite drastically over time and geographic position, which requires an increasing flexibility of the power system~\cite{energy_graabak,energy_huber}. Efficient ways of storing energy can increase the flexibility of the power system~\cite{storage_gur,storage_lai,storage_matos,storage_pasada}, allowing us to successfully make the transition towards renewable energy.

Technologies for  energy storage vary immensely, and several classifications can be considered, for instance the actual technology used (e.g. chemical, electrochemical, electrical, mechanical and thermal) or the scale of storage (e.g. small scale, short term, large scale, long term); each of these subcategories has its own strengths, weaknesses and range of applicability~\cite{storage_bullich,storage_matos}. One of these technologies employs electric double layers (EDLs), in which the spatial distribution of mobile charges of electrolytes in contact with electrodes allows for storage of electric energy. Basically, such a device is comprised of a charged conductor (an electrode) immersed in an electrolyte, where the electron charge on the electrode is screened by the ions in the electrolyte. Such a system behaves in some sense similar to a conventional capacitor of two electrodes, however with a much smaller distance between the oppositely charged entities, which in this case consists of an immobile electrode with, say, a positive charge in close contact with a layer of mobile ions in the electrolyte with a net negative charge. Generally, the electrodes of EDL capacitors have a porous structure with a gigantic surface-to-volume ratio (up to $2000$ m$^2$ g$^{-1}$)~\cite{Gorka_2010}, and no chemical reactions take place at the surface. The large area leads to a large capacitance and therefore a large amount of energy stored compared to conventional capacitors and the absence of reactions leads to long lifetimes compared to conventional batteries, which makes them ``greener" as well. One of the disadvantages of EDL capacitors is the rather small energy density compared to common Lithium-ion batteries (by roughly a factor of 20)~\cite{storage_gur}.

Within a density functional theory that captures Coulomb and hard-sphere interactions accurately~\cite{Cats_2020}, we calculate the differential capacitance for several electrolytes, changing the size and valency of the ions similar to work done in Refs.~\onlinecite{Bossa_2019,Kaja_2016,Henderson_2017,Biesheuvel_2017,Caetano_2016,Nakayama,Kornyshev_2007,Frischknecht_2014,Ma_2014,Vo_2020}. However, we derive a new relation for the differential capacitance, which explicitly shows that the response of the first layer of ions near the electrode to the applied potential on the electrode is key to its understanding. We utilize this insight to understand two- and three-component electrolytes and explain how impurities can have a large effect on the differential capacitance.

\section{\label{sec:level1}Differential Capacitance}

\subsection{The System}
\begin{figure*}
    \centering
    \includegraphics[width=0.9\textwidth]{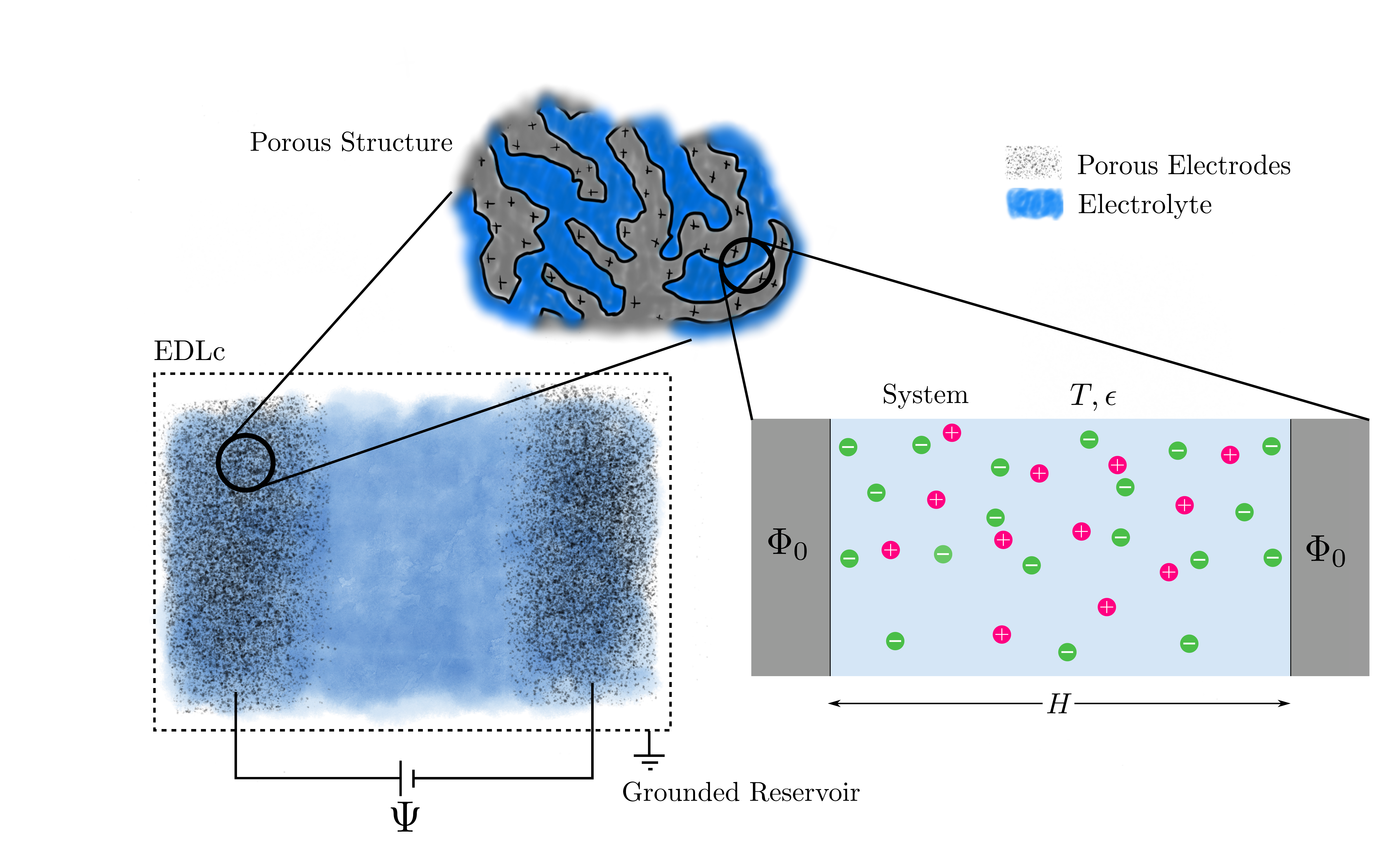}
    \caption{An illustration of an EDL capacitor and its structure. The EDL capacitor within the dashed box contains two porous electrodes (black dots), held at a potential difference $\Psi$, through which the electrolyte (blue) can flow.  Zooming in on a porous electrode reveals a lot of structure which is depicted on top. Throughout this manuscript the focus lies on one tiny part of the electrode (the pore of interest), which we consider our system, and model it as two parallel charged surfaces separated by a distance $H$ being part of the same electrode. Hence, the potential on the two surfaces, denoted by $\Phi_0$, is the same and measured w.r.t. the potential in the reservoir, which we assume to be grounded. Invoking the primitive model reduces the solvent to a continuous dielectric medium at temperature $T$ and with uniform dielectric constant $\epsilon$.} 
    \label{Fig:pore_system}
\end{figure*}

As mentioned in the introduction, we wish to focus on EDL capacitors, of which there is a variety of different types, each having a different electrode-electrolyte combination~\cite{EDLc_Chen}. For now, we consider two porous (carbon) electrodes held at a potential difference $\Psi$ immersed in an aqueous electrolyte at room temperature $T$ and dielectric constant $\epsilon$. We will focus on a single pore inside one electrode, and model it as two planar surfaces with the same static surface potential $\Phi_0$, separated by a distance $H$~\cite{EDLCs_zhai,Winter_2004}, as depicted in Fig.~\ref{Fig:pore_system}. We stress that the surface potential $\Phi_0$ on the surfaces of the pore is w.r.t. a potential in a macroscopic electrolyte reservoir (the space between the two electrodes), while $\Psi$ is the potential difference between the two electrodes in the EDL capacitor.

We will employ the primitive model (PM) in which the solvent is treated as a continuous dielectric medium. The properties of the solvent can then be completely captured by a single parameter: the Bjerrum length $\lambda_B=e^2\beta/4\pi\eps$, where $e$ is the proton charge and $\beta=1/k_BT$ the inverse temperature. The Bjerrum length of water at room temperature is $\lambda_B=0.73$ nm, which will be kept constant throughout this work. The ions of species $j$ are modelled as hard spheres with diameter $d_j$ and a centro-symmetric charge $ez_j$ with $z_j$ denoting the valency; the pair potential $u_{ij}(r)$ between ions of species $i$ and $j$ separated by a distance $r$ is described by
\begin{align}
    \beta u_{ij}(r)=\begin{cases}
     \infty, \quad &r<d_{ij};\\
     z_iz_j\frac{\lambda_B}{r}, \quad &r\geq d_{ij},
    \end{cases}
\end{align}
where $d_{ij}=(d_i+d_j)/2$. Note that the PM neglects dispersion and polarizability. We denote the bulk concentration of ion species $j$ in the macroscopic electrolyte reservoir between the two electrodes by $\rho_{b,j}$, and define the total ionic packing fraction in the bulk as
\begin{align}\label{Eq:eta_bulk}
    \eta_b=\frac{\pi}{6}\sum_j d_j^3\rho_{b,j}.
\end{align}
When all particles have the same diameter and valency, the model is referred to as the Restrictive Primitive Model (RPM). The bulk RPM is completely characterized by the ionic blk concentration and the dimensionless temperature $T^*=d/\lambda_B$. We stress that the medium and the diameters are chosen such that $T^*\gg T_c^*=0.05$ the critical temperature~\cite{Orkoulas_1994}, e.g. in our manuscript $\lambda_B=0.73$ nm is fixed and the ion diameters range from $d_j=0.2$ nm to $d_j=1$ nm, for which $T^*$ ranges from $T^*=0.27$ to $T^*=1.37$, respectively.

The (non-electrostatic part of the) external potential that determines the pore size reads
    \begin{align}\label{Eq:Vext}
    V_{ext}^j(x)=\begin{cases}
     \infty \quad x\leq\frac{d_j}{2};\\
     0\quad \frac{d_j}{2}<x<H-\frac{d_j}{2};\\
     \infty \quad x\geq H-\frac{d_j}{2},\\
    \end{cases}
\end{align}
which describes two planar hard walls positioned at $x=0$ and $x=H$. Throughout this manuscript, the surface separation is fixed at $H=5$ nm. Note that the electrostatic part of the external potential, due to the surface charge on the electrodes, will be taken into account through the Poisson equation, as discussed in Appendix~\ref{App:MSA}.

Within this simplification, we can controllably study the differential capacitance as a function of the surface potential $\Phi_0$, for several choices of the ion diameter $d_j$, valency $z_j$, and ionic bulk concentrations in the reservoir $\rho_{b,j}$.

\subsection{Thermodynamics in the Pore}

Let us now focus on a single pore as depicted in Fig.~\ref{Fig:pore_system}. The remainder of the electrode and the space between the electrodes we consider to be the ion and heat reservoir for the pore of interest. Assuming that the space between the electrodes is large enough, we can treat the pore of interest grand canonically. This justifies the introduction of a bulk ionic concentration $\rho_{b,j}$ (corresponding to a fixed ionic chemical potential $\mu_j$) in the reservoir that is independent of the electric potential difference $\Psi$ between the electrodes. Exchange of ions between the system and reservoir takes place and we assume equilibrium. The surface potential $\Phi_0$ on the surfaces in the pore is fixed, which allows exchange of electronic charge between the electrode and a charge reservoir; however, no charge exchange takes place between electrode and ions, i.e. no electrochemistry is considered. The pore also has a fixed volume $V=AH/2$ (rigid electrode) with total surface area $A$, and temperature $T$. To summarize, the system is described by a $\{\mu\},V,T,\Phi_0,A,H$ ensemble, characterized by the grand potential $\Omega(\{\mu\},V,T,\Phi_0,A,H)$ ~\cite{Cats_2020}, where the brackets $\{\mu\}$ denote the set of chemical potentials of each of the ion species. The differential of $\Omega$ for this system reads 
\begin{align}\label{Eq:dOm}
    \dd \Omega=&-p\dd V-S\dd T-\sum_jN_j\dd\mu_j +\\
    &-Q\dd\Phi_0 + \gamma \dd A - f\dd H,\nonumber
\end{align}
where $N_j$ is the average number of ions of species $j$ in the pore, $p$ the pressure, $Q$ the electronic charge on both surfaces of the pore, $\gamma$ the surface tension,  and $f$ the force between the surfaces that are separated by a distance $H$. It is convenient to separate the volumetric and areal contribitions in Eq.~\eqref{Eq:dOm}, and write $\Omega=-p(T,\{\mu\})V+\gamma(T,\{\mu\},\Phi_0,H)A$  as two separate differentials~\cite{Cats_2020}
\begin{align}
    \dd p&=s_b\dd T+\sum_j\rho_{b,j} \dd\mu_j,\label{Eq:dp}\\
    \dd\gamma&=-s_s\dd T-\sum_j\Gamma_j\dd\mu_j-\sigma\dd\Phi_0-\frac{f}{A}\dd H,\label{Eq:dgam}
\end{align}
where $S$ and $N_j$ have been separated into a volumetric and areal part according to $S=Vs_b+As_s$, and  $N_j=V\rho_{b,j}+A\Gamma_j$ with $\rho_{b,j}$ the bulk concentration of species $j$ and $\Gamma_j$ the adsorption. The average electronic surface charge density is denoted by $\sigma=Q/A$ and obeys the charge neutrality condition 
\begin{align}
    \sigma=-\sum_jez_j\Gamma_j.
\end{align}
Eq.~\eqref{Eq:dp} is the Gibbs-Duhem equation and Eq.~\eqref{Eq:dgam} the Lipmann equation.

Now the question arises: what does this have to do with capacitances? The short answer to this question is that the (areal) differential capacitance is defined by
\begin{align}
    C_{\mu,T,H}&=\left(\frac{\partial \sigma}{\partial \Phi_0}\right)_{\mu,T,H}\label{Eq:Cmu}\\
    &=-\left(\frac{\partial^2\gamma}{\partial \Phi_0^2}\right)_{\mu,T,H},\label{Eq:Cmu_Omega}
\end{align}
which opens the door for the longer answer. First of all, it turns out that there is not \textit{one} differential capacitance $C$, but a whole set of differential capacitances, one for each ensemble. This is similar to constant-volume and constant-pressure heat capacities, and isothermal and isentropic compressibilities, for instance. Throughout this manuscript we will study the differential capacitance at fixed temperature $T$ and surface separation $H$, and simplify the notation: the ensemble in Eq.~\eqref{Eq:dOm} gives rise to $C_\mu$, while an $\{N\},V,T,\Psi,H$ ensemble would give rise to $C_N$. Those two are, in fact, not  unrelated and follow similar thermodynamic identities as those of the heat capacities at constant volume or pressure~\cite{Roijstat,Cats_2020}. We will focus on $C_\mu(\Phi_0,\{\mu\})$.

\subsection{Differential Capacitance Revisited}

Before tackling the primitive-model electrolyte confined between two hard walls by Density Functional Theory (DFT), let us first discuss the mechanical equilibrium in a planar geometry. The $xx$ (normal) component of the pressure tensor within the pore follows from the force balance~\cite{Henderson_1978,Henderson_1979}
\begin{align}\label{Eq:sum_rule_general}
    p_{xx}(H)=&\sum_j \int_0^H \dd x \rho_j(x)F_x^j(x),
\end{align}
where $\rho_j(x)$ is the local density of species $j$, $F_x^j(x)=-\partial V_j(x)/\partial x$ the normal force on the surface due to the full interaction potential $V_j(x)$ (including the electrostatic interactions due to the surface charge).
If the integral in Eq.~\eqref{Eq:sum_rule_general} does not depend on $H$, which is the case when the range of $F_x^j$ is much smaller than $H$, then $p_{xx}(H)=p$ is the actual bulk pressure of the electrolyte and Eq.~\eqref{Eq:sum_rule_general} reduces to~\cite{Henderson_1978,Henderson_1979}
\begin{align}\label{Eq:p_sumrule}
    \beta p=\sum_j \rho_j\left(\frac{d_j^+}{2}\right)-\frac{\beta \sigma^2}{2\eps},
\end{align}
where $d_j^+=\lim_{x\downarrow d_j}x$.
    This sum rule relates the bulk pressure $p$ of the ion reservoir to the local contact densities $\rho_j(d_j^+/2)$ and the surface charge density $\sigma$. In the case of two charged hard parallel surfaces separated by a finite distance, the integral in Eq.~\eqref{Eq:sum_rule_general} does depend on $H$, especially at low ionic bulk concentrations. However, with the parameters chosen throughout this manuscript, this only plays a minor role and can be neglected for all practical purposes. Let us note that for soft particle-wall interactions the integral in Eq.~\eqref{Eq:sum_rule_general} does not reduce to a contact value and needs to be fully evaluated. Nonetheless, the gradient of the particle-wall interaction is by construction largest at values close to the contact distance $d_j/2$ at which it gives the largest contribution to the integral.
    
    One interesting feature of Eq.~\eqref{Eq:p_sumrule} is that the global bulk pressure is independent of the surface potential $\Phi_0$, while the local quantities are not.   Therefore, taking the derivative of Eq.~\eqref{Eq:p_sumrule} w.r.t. $\Phi_0$ leads to
\begin{align}\label{Eq:dpdphi}
 0=   \sum_j\left(\frac{\partial\rho_j(d_j^+/2)}{\partial \Phi_0}\right)_\mu -\frac{\beta \sigma}{\eps}\left(\frac{\partial \sigma}{\partial\Phi_0}\right)_\mu,
\end{align}
where we recognize the differential capacitance given in Eq.~\eqref{Eq:Cmu} in the second term. Hence, Eq.\eqref{Eq:dpdphi} can be rewritten as
\begin{align}\label{Eq:Cap_sumrule}
C_\mu&=\frac{\eps}{\beta\sigma}\sum_j\left(\frac{\partial\rho_j(d_j/2)}{\partial \Phi_0}\right)_\mu,
\end{align}
where in the summand one can recognize a fluctuation profile as discussed in Ref.~\onlinecite{Eckert_2020}.
Within the RPM, Eq.~\eqref{Eq:Cap_sumrule} simplifies to
\begin{align}\label{Eq:Cap_sumrule_RPM}
C_\mu=\frac{\eps}{\beta\sigma}\left(\frac{\partial\rho_N(d/2)}{\partial \Phi_0}\right)_\mu,
\end{align}
where we introduce the total ion density
\begin{align}
    \rho_N(x)=\sum_j\rho_j(x).
\end{align}
Hence, the differential capacitance is to a large extent dictated by the response of the number density $\rho_N(d/2)$ at contact (first layer of ions) to the potential $\Phi_0$. As stated previously, this key-concept will stand even for soft particle-wall interactions; the contact value will be replaced by an integral, whose integrand is peaked near $x=d_j/2$.

Interestingly, another observation leads to yet another expression for the differential capacitance. Due to the ions having a hard core interaction with the wall and a central charge, the charge density for $x<d/2$ (also for $x>H-d/2$) vanishes, leading to a linear drop of the electrostatic potential profile $\Phi(x)$ in that region. Hence, the potential at $x=d/2$ is given by
\begin{align}
    \Phi(d/2)=\Phi_0+\left.\frac{\partial \Phi(x)}{\partial x}\right|_{x=0}\frac{d}{2}=\Phi_0-\frac{d}{2\eps}\sigma,
\end{align}
where we used Gauss' law.
Taking again the derivative w.r.t. the surface potential $\Phi_0$ and reordering results in
\begin{align}\label{Eq:C_phi}
   C_\mu=\frac{2\eps}{d}\left[1 -\left(\frac{\partial \Phi(d/2)}{\partial \Phi_0}\right)_\mu\right].
\end{align}
The region between $x=0$ and $x=d/2$ is sometimes referred to as the  Stern layer~\cite{Stern}, with the corresponding Stern-layer capacitance~\cite{Kornyshev_2007,vanhal_1996,Locket_2010,May_2019} $C_s=2\eps/d$, although that will not be used in this work~\footnote{The Stern layer  $C_{EDL}$ separates the diffuse layer from the electrode, each layer having their own capacitance, which in a serial circuit gives the total capacitance of the EDL:
\begin{align}\label{Eq:C_series_EDL}
    \frac{1}{C_{EDL}}=\frac{1}{C_s}+\frac{1}{C_{DL}}.
\end{align}
With little effort, one can show using Eqs.~\eqref{Eq:C_phi} and~\eqref{Eq:C_series_EDL} that, by introducing the Stern-layer capacitance, the capacitance of the diffuse layer reads
\begin{align}
    C_{DL}=C_s\left[\left(\frac{\partial \Phi_0}{\partial \Phi(d/2)}\right)_\mu-1\right],
\end{align}
which can take negative values in case of overscreening. Hence, splitting the EDL into a diffuse layer and the stern layer seems not to be desirable for a generic system, and we will not further distinguish between Stern layers and other layers, but instead talk about the total capacitance $C_\mu$.}.

The relations in Eqs.~\eqref{Eq:Cap_sumrule} and~\eqref{Eq:C_phi} will allow for a new and independent investigation into the differential capacitance, which helps in understanding the underlying physics.

\subsection{Density Functional Theory}
We will use classical DFT to study the PM electrolytes. We refer to previous literature for the details~\cite{Evans_1979,Roth_2010,Haertel_2017}. In short, DFT involves a grand potential functional, which reads
\begin{align}
    \Omega[\{\rho\}]=&\F_{id}[\{\rho\}]+\F_{ex}^{HS}[\{\rho\}]+\F_{ex}^{ES}[\{\rho\}]-\nonumber\\
    &\sum_j\int \dd \rb \rho_j(\rb)\left(\mu_j - V_{ext}^j(\rb)\right)-Q\Phi_0,
\end{align}
where $V_{ext}^j$ is the (non-electric) external potential of Eq.~\eqref{Eq:Vext} acting on particles of species $j$, and $Q=\sigma A$ is the total charge on the surface in the pore. The first term $\F_{id}[\{\rho\}]$ is the Helmoltz free energy functional for an ideal (non-interacting) system, the second term $\F_{ex}^{HS}[\{\rho\}]$ accounts for the hard-sphere interactions and is dealt with by Fundamental Measure Theory White-Bear II~\cite{Roth_2010}, and the third term $\F_{ex}^{ES}[\{\rho\}]$ describes the Coulombic interactions for which we invoke the MSAc functional described extensively in Refs.~\onlinecite{MSAc,Cats_2020_decay} (see also Appendix~\ref{App:MSA}). 

The grand potential functional $\Omega[\{\rho\}]$ is, in fact, minimized by the equilibrium density profiles $\{\rho_0\}$~\cite{Evans_1979}, i.e.
\begin{align}
    \left.\frac{\delta \Omega[\{\rho\}]}{\delta \rho_j(\rb)}\right|_{\{\rho\}=\{\rho_0\}}=0.
\end{align}
Moreover, the grand potential at its minimum is the actual grand potential of the system, i.e. $\Omega[\{\rho_0\}]=\Omega(\{\mu\},V,T,\Phi_0,H)$. Hence,  DFT is a powerful theoretical framework to combine thermodynamics and structure as it gives access to the thermodynamics of the system via the density profiles. Furthermore, it gives directly both the ionic charge density profile $q(x)=\sum_j z_j \rho_j(x)$ and via the Poisson equation also the potential profile $\Phi(x)$ from which the surface charge follows, where we note that in a planar geometry the profiles only depend on the out-of-plane coordinate $x$.  In principle, we have thus access to $C_\mu$ via any of the routes in Eqs.~\eqref{Eq:Cmu},~\eqref{Eq:Cmu_Omega},~\eqref{Eq:Cap_sumrule}, and~\eqref{Eq:C_phi} laid out in the previous sections; however, the results that will be presented are calculated from Eq.~\eqref{Eq:Cmu}, while the other definitions are used as a tool to interpret the data. Nevertheless, we have validated each of the Eqs.~\eqref{Eq:Cmu},~\eqref{Eq:Cmu_Omega},~\eqref{Eq:Cap_sumrule}, and~\eqref{Eq:C_phi} for the sake of consistency.

\subsection{Simulations}
Although DFT is a very powerful framework, the intrinsic functionals $\F_{ex}^{HS}$ and $\F_{ex}^{ES}$ are approximate and need to be tested against simulations or experiments. Although we do not go into detail here on any specific type of simulation, we wish to present a novel sumrule that connects quantities that can be measured in simulation but not with DFT. The quantity of interest throughout this manuscript is the differential capacitance. From elementary statistical physics one can derive the relation~\cite{Limmer_2013,Scalfi_2020,Roijstat}
\begin{align}\label{Eq:C_fluct}
    C(\Phi_0)=\frac{\beta}{A}\left(\langle Q^2\rangle - \langle Q \rangle^2\right),
\end{align}
where the brackets $\langle\ldots\rangle$ indicate either a time or an ensemble average.

Inspired by Ref.~\onlinecite{Eckert_2020}, we also found a new expression to calculate the differential capacitance, namely 
\begin{align}
C&=\frac{\eps}{\sigma}\sum_j\left[ \langle \hat{\rho}_j(d_j/2) Q \rangle-\langle\hat{\rho}_j(d_j/2)\rangle\langle  Q \rangle\right],\label{Eq:C_fluct_prof}
\end{align}
where $\hat{\rho}_j(\rb)=\sum_{i=1}^{N_j} \delta(\rb-\rb_i)$ denotes the density operator.
Note that we use the notation $\sigma=\langle\sigma\rangle$ throughout the manuscript, but it is important to keep in mind that the surface charge $\sigma=Q/A$ fluctuates as a consequence of fixing the surface potential $\Phi_0$. 

Employing DFT causes the loss of direct information on the fluctuations, because DFT only returns equilibrium density profiles. Hence, within our DFT approach one does not have direct access to the quantities such as $\langle Q^2 \rangle$ and $\langle  \hat{\rho}_j(d_j/2) Q \rangle$. However, they can in principle be measured within simulations.

\section{Capacitance Curves}\label{Sec:Results}

\subsection{First Glance: Symmetric Electrolyte}
\begin{figure}
    \centering
    \includegraphics[width=\columnwidth]{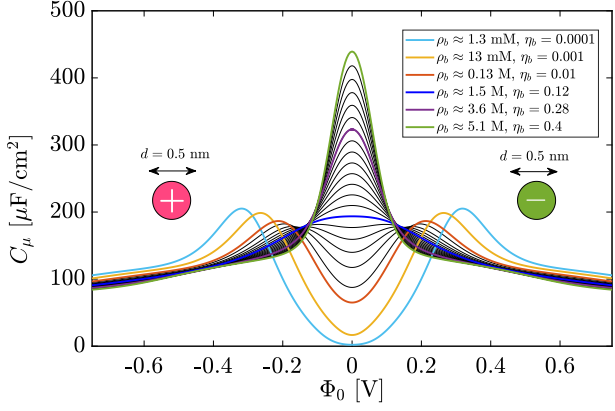}
    \caption{The differential capacitance $C_\mu$ for the RPM with monovalent ions with diameter $d=0.5$ nm and for a surface separation $H=5$ nm and Bjerrum length $\lambda_B=0.73$ nm. The curves are shown for concentrations ranging from $\rho_b=1.3$ mM  (light-blue) to $\rho_b=5.1$ M (green) with several concentrations in between. The concentrations for the other colored lines are given in the legend; the concentration for the black lines are equally spaced to illustrate the trend. The legend also shows the corresponding bulk packing fractions $\eta_b$.}
    \label{Fig:cap_RPM}
\end{figure}
The system that we investigate first is an RPM electrolyte consisting of two monovalent species of ions (i.e. $z_+=-z_-=1$), so that $\rho_{b,+}=\rho_{b,-}\equiv\rho_b$. In Fig.~\ref{Fig:cap_RPM} we show the differential capacitance, calculated via Eq.~\eqref{Eq:Cmu}, as a function of the applied voltage $\Phi_0$ for ions with $d=0.5$ nm at several ionic bulk packing fractions ranging from $\eta_b=0.0001$ to $\eta_b=0.4$, corresponding to bulk ionic concentrations ranging from $\rho_b\approx 1$ mM to $\rho_b\approx5$ M. As stated previously, the surface separation is fixed at $H=5$ nm and the solvent is characterized by $\lambda_B=0.73$ nm. 
The foremost noticeable feature of Fig.~\ref{Fig:cap_RPM} is the crossover around a bulk packing fraction $\eta_c\approx 0.12$ (blue line) from the so-called camel-shaped ($\eta_b<\eta_c$) to the bell-shaped curves ($\eta_b>\eta_c$). This crossover has been a subject in many theoretical and experimental studies~\cite{Kornyshev_2007,Fedorov_2010,Fedorov_2014,Uematsu_2018,Girotto_2018,Dowing_2018,Islam_2008,Jiang_2011,Nakayama,Jitvisate_2018}, and is ascribed to excluded-volume effects. Let us, for convenience, introduce $\Phi^*$ as the surface potential at which the capacitance $C_\mu(\Phi_0)$ takes its maximum value at a given $\rho_b$ or $\eta_b$, i.e. $C_\mu(\Phi^*)\geq C_\mu(\Phi_0)$. Then the camel-shaped capacitance curves correspond to a finite $\Phi^*$, while the bell-shaped capacitance curves are characterized by a vanishing $\Phi^*$.

\begin{figure}
    \centering
    \includegraphics[width=\columnwidth]{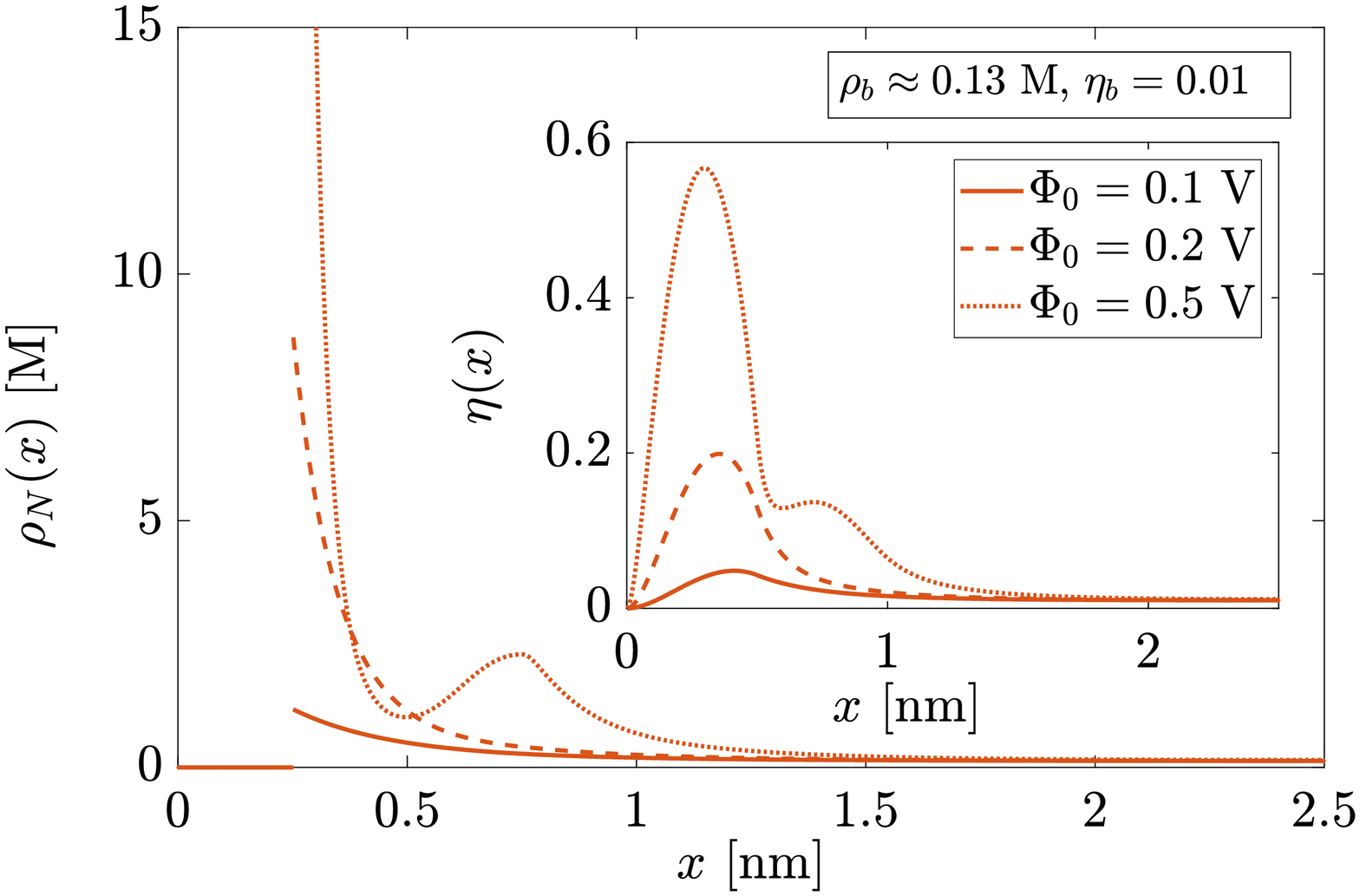}
         \includegraphics[width=\columnwidth]{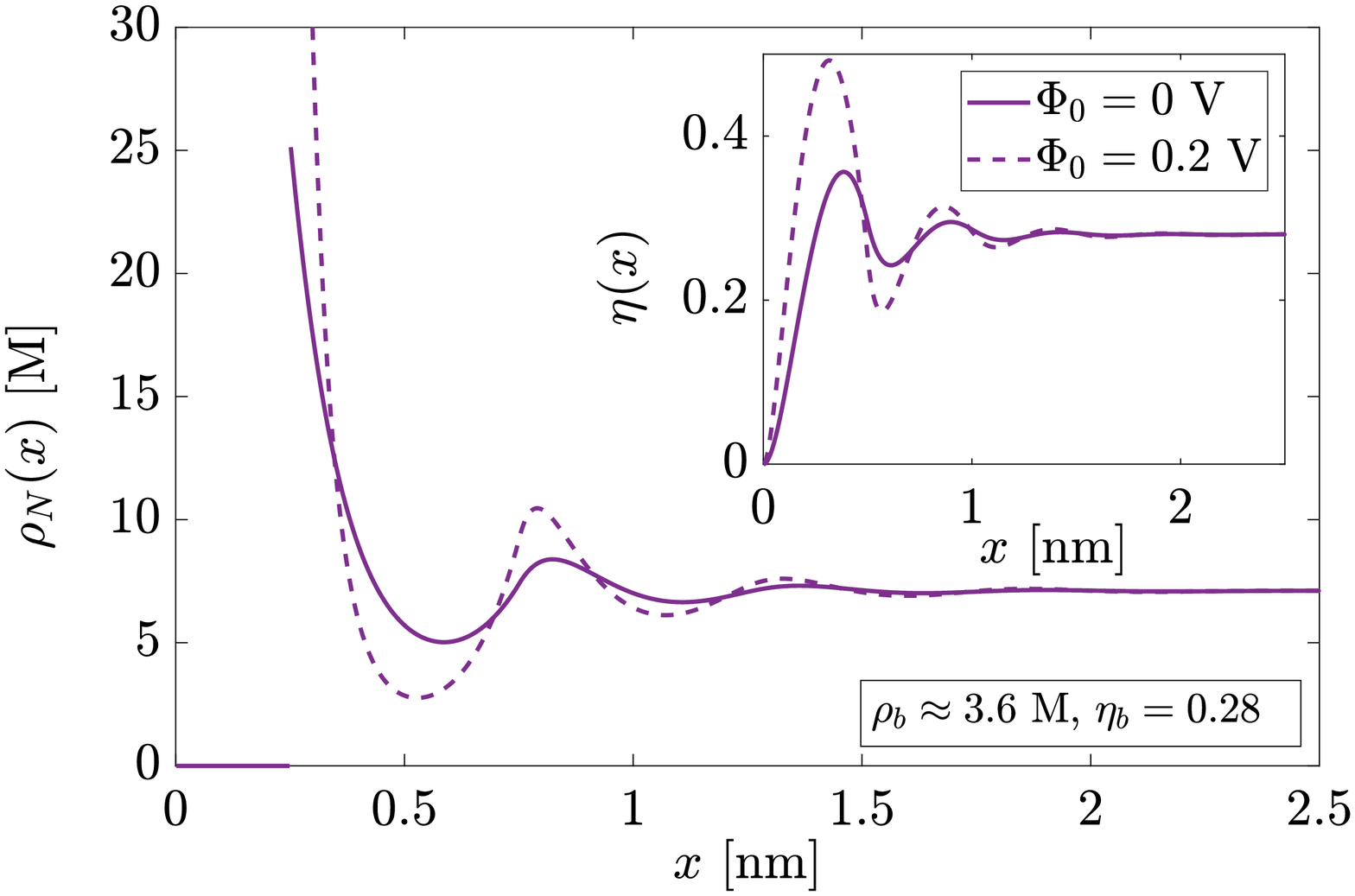}
    \caption{(a) The number density profiles $\rho_N(x)$ for the surface potentials $\Phi_0=0.1$ V (solid), $\Phi_0=0.2$ V (dashed), and $\Phi_0=0.5$ V (dotted) at a bulk concentration of $\rho_b\approx 0.13$ M (i.e. $\eta_b=0.01$). The inset shows the local packing fraction $\eta(x)$ as defined in Eq.~\eqref{Eq:eta_local}. The potential at which the capacitance has its maximum is $\Phi^*\approx 0.2$ V. For $\Phi_0<\Phi^*$ the profiles only have one peak at $x=d/2=0.25$ nm, while for $\Phi_0>\Phi^*$ the profiles have two peaks; the second located at $x=3d/2=0.75$ nm. Around the maximum at $\Phi_0\approx \Phi^*$ the second layer starts forming.
    (b) The number density profiles $\rho_N(x)$ for the surface potentials $\Phi_0=0$ V (solid) and $\Phi_0=0.2$ V (dashed) at a bulk concentration of $\rho_b\approx 3.6$ M (i.e. $\eta_b=0.28$). The inset shows the local packing fraction $\eta(x)$ as defined in Eq.~\eqref{Eq:eta_local}.  This bulk concentration $\rho_b\approx 3.6$ M corresponds to the same-colored (purple) bell-shaped capacitance curve in Fig.~\ref{Fig:cap_RPM}. }
    \label{Fig:rho_x_eta}
\end{figure}

To understand this crossover, one has to understand what causes the maximum in $C_\mu(\Phi_0)$.  Consider the orange camel-shaped capacitance curve in Fig.~\ref{Fig:cap_RPM} belonging to $\rho_b\approx 0.13$ M (i.e. $\eta_b=0.01<\eta_c$) with $\Phi^*\approx0.2$ V. For the same $\rho_b$ (and in the same color) we present in Fig.~\ref{Fig:rho_x_eta}(a) the number density profiles $\rho_N(x)$ for $\Phi_0=0.1$ V (solid), $\Phi_0=0.2$ V (dashed), and $\Phi_0=0.5$ V (dotted). The inset shows, for the same state points, the profile of the weighted packing fraction, defined by \footnote{$\eta(x)$ is identical to the weighted density $n_3(x)$ from Fundamental Measure Theory~\cite{Roth_2010}.}
 \begin{align}\label{Eq:eta_local}
     \eta(x)=\pi\sum_j \int_{x-d_j/2}^{x+d_j/2}\dd x' \rho_j(x')\left[ \left(\frac{d_j}{2}\right)^2-(x-x')^2\right],
 \end{align}
which reduces to Eq.~\eqref{Eq:eta_bulk} in the homogeneous bulk. Interestingly, while the density profile for $\Phi_0<\Phi^*$ (solid) only has one (contact) peak at $x=d/2=0.25$ nm, the one for $\Phi_0>\Phi^*$ (dotted) has a second peak at $x=3d/2=0.75$ nm. Hence, the maximum in $C_\mu$ indicates a structural change, analogous to peaks in the heat capacity indicating (smooth) changes of the thermal occupancy of microstates. When we consider the purple bell-shaped capacitance curve in Fig.~\ref{Fig:cap_RPM}, belonging to $\rho_b\approx 3.6$ M (i.e. $\eta_b=0.28>\eta_c$), and investigate the corresponding $\rho_N(x)$ and $\eta(x)$ in Fig.~\ref{Fig:rho_x_eta}(b) for $\Phi_0=0$ V (solid) and $\Phi_0=0.2$ V (dashed), we find the presence of the second peak in $\rho_N(x)$ for both (and in fact for all) surface potentials. Hence, the maximum in $C_\mu(\Phi_0)$ is clearly related to the onset of layering of counter-ions near the surface, as is in fact  also consistent with Eq.~\eqref{Eq:Cap_sumrule}. Moreover, considering the profile of the weighted packing fraction (see insets in Figs.~\ref{Fig:rho_x_eta}(a) and (b)) at $x=d/2=0.25$ nm, one finds that its value is close to or even larger than the packing fraction at which hard-sphere freezing takes place in the bulk (i.e. around $\eta_b\approx0.5$). This suggests that the first layer of ions might in fact even be frozen~\cite{Dijkstra_2004}, and invites a study into the in-plane structure of ions at the charged surface. However, that is beyond the scope of this manuscript.

\begin{figure}
     \centering
     \includegraphics[width=\columnwidth]{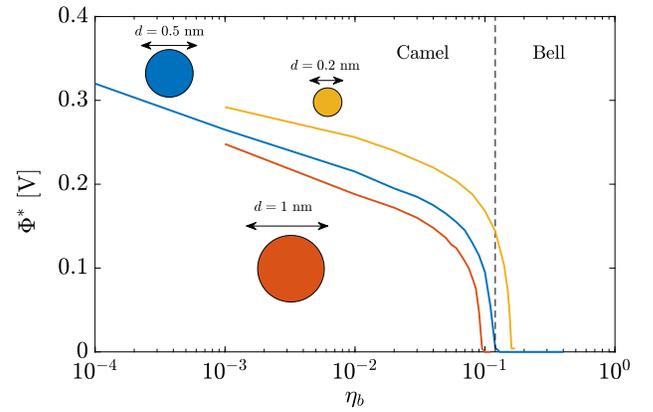}
         \caption{The surface potential $\Phi^*$ at which $C_\mu(\Phi_0)$ has its maximum as function of the bulk concentration expressed as the bulk packing fraction $\eta_b$, for the RPM with ion diameter $d=0.2$ nm (yellow), $d=0.5$ nm (blue), and $d=1$ nm (orange). The vertical dashed line is located at $\eta_b=0.12$ and represents the crossover for ions with $d=0.5$ nm from camel-shaped capacitance curves to bell-shaped capacitance curves.}
     \label{Fig:etab_Phi0s_R_25}
 \end{figure}

To characterize and interpret the crossover from camel- to bell-shaped curves further, let us consider the crossover potential $\Phi^*(\eta_b)$ as function of the bulk packing fraction $\eta_b$ plotted in Fig.~\ref{Fig:etab_Phi0s_R_25}, for three ion diameters. The dashed line at $\eta_c\approx 0.12$ represents for $d=0.5$ nm the capacitance curve crossing over from two maxima at $\Phi_0=\Phi^*>0$ (camel) to one maximum at $\Phi_0=\Phi^*=0$ (bell). From Fig.~\ref{Fig:etab_Phi0s_R_25}, it is evident that the crossover is rather gradual and therefore we refer to a crossover rather than a transition. This is in line with the maximum in the capacitance curves being finite rather than infinite, the latter case would imply a thermodynamic phase transition.
 This graduality is found in the density profiles as well: where profiles for $\Phi_0\ll \Phi^*$ and $\Phi_0\gg \Phi^*$ can be easily discerned by the absence and presence of a second peak at $x=1.5d=0.75$ nm (see Fig.~\ref{Fig:rho_x_eta}(a)), those for potentials close to $\Phi^*$ cannot. The same holds for density profiles at $\Phi_0=0$; for low bulk concentrations layering is almost absent, while at high bulk concentrations layering is clearly present (see Fig.~\ref{Fig:rho_x_eta}(b)), but one cannot qualitatively discern the two cases close to $\eta_b\approx \eta_c$.\\

\begin{figure}
     \centering
     \includegraphics[width=\columnwidth]{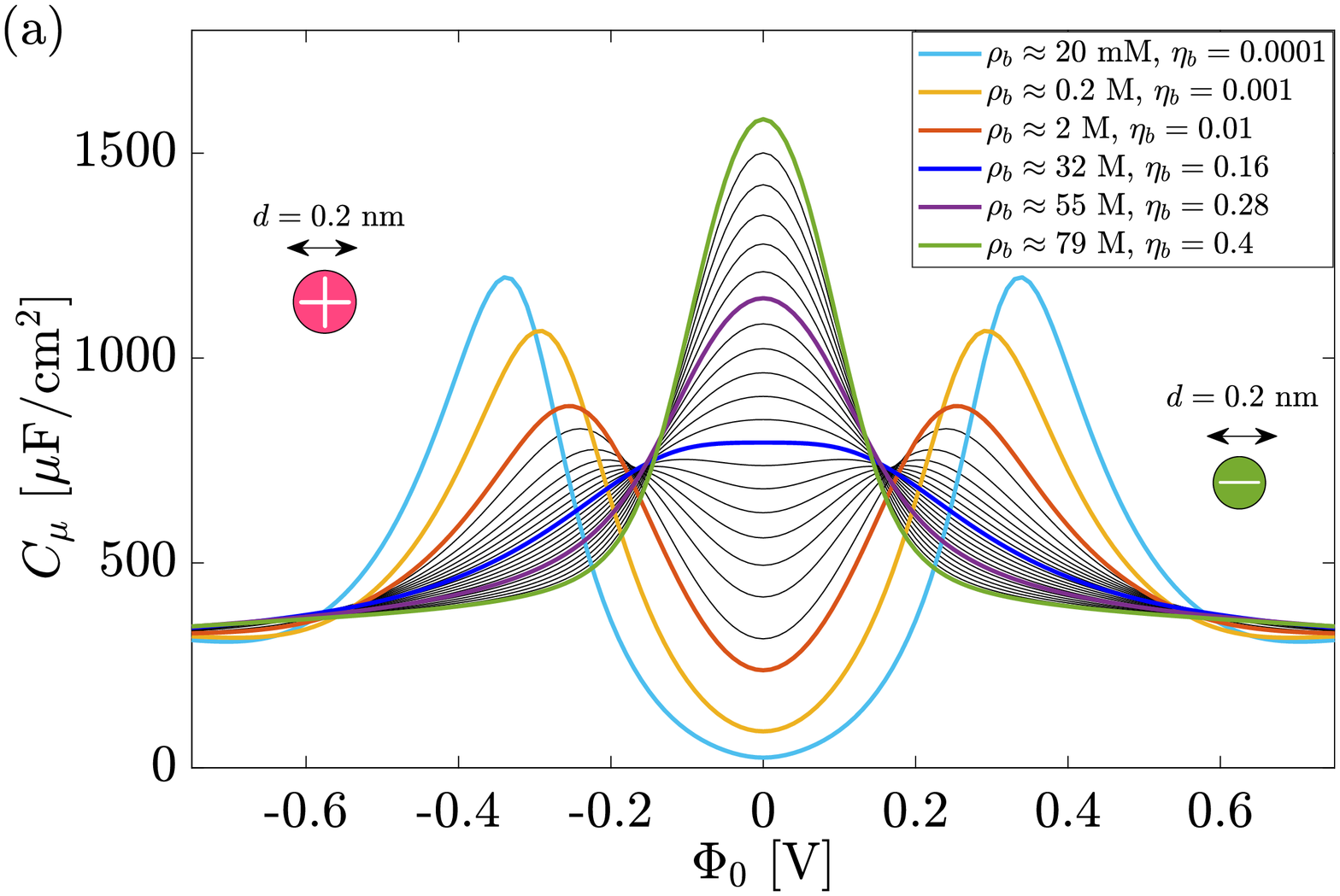}
     \includegraphics[width=\columnwidth]{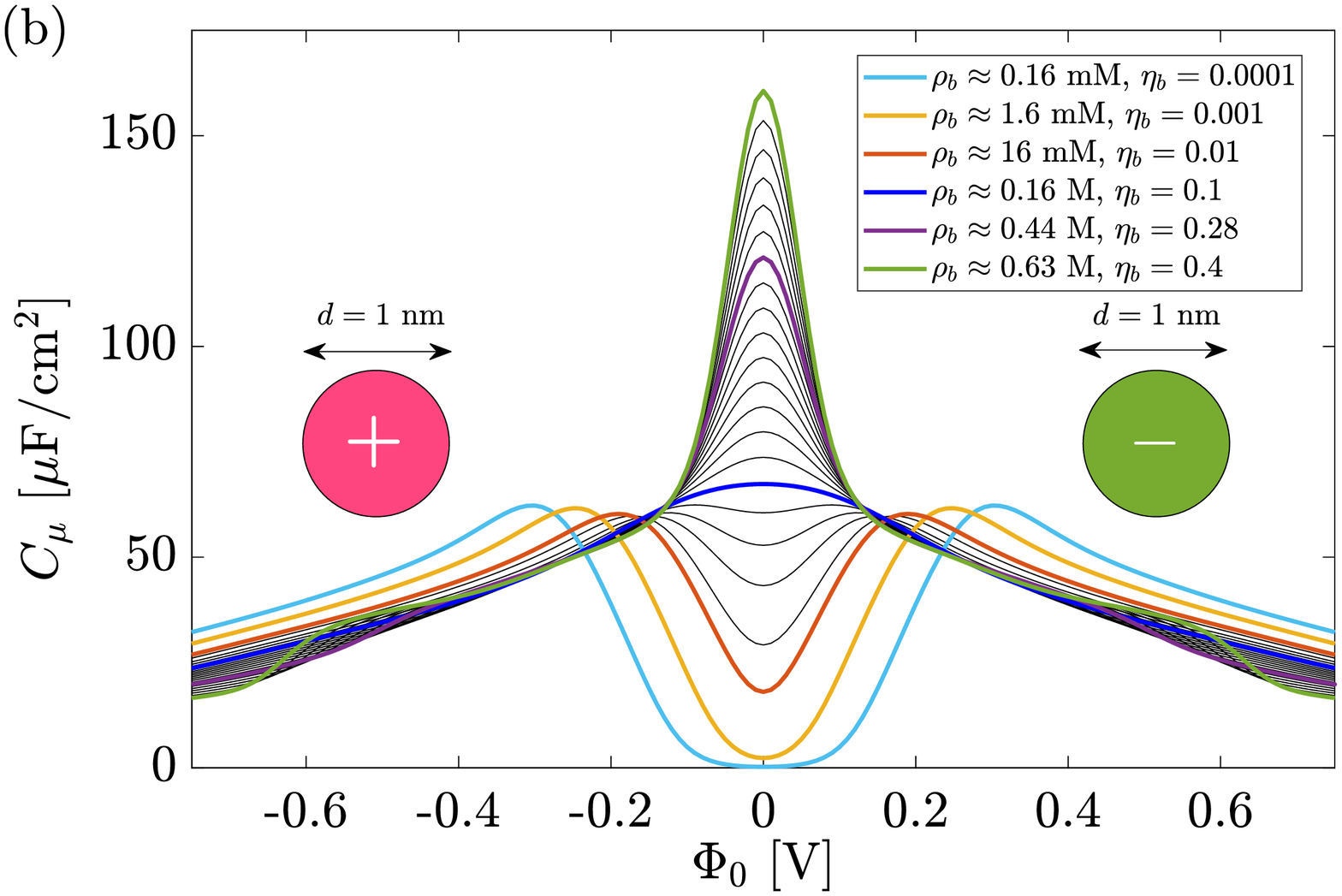}
         \caption{The differential capacitance for ions with diameter (a) $d=0.2$ nm and (b) $d=1$ nm for a surface separation $H=5$ nm and Bjerrum length $\lambda_B=0.73$ nm. The bulk concentration/packing fraction at which the camel-bell crossover occurs is depicted in dark blue, i.e. $\eta_c\approx 0.1$ ($\rho_c\approx 32$ M) for $d=0.2$ nm and $\eta_c\approx 0.16$ ($\rho_c\approx 0.16$ M) for $d=1$ nm. }
     \label{Fig:C_R_10_50}
 \end{figure}

 The results so far were all for ions with a common diameter $d=0.5$ nm and $\lambda_B=0.73$ nm. Let us now investigate the effect of ion size, keeping $\lambda_B$ fixed at $\lambda_B=0.73$ nm. To this end, we next consider ions of a common diameter $d=0.2$ nm and $d=1$ nm, for which $C_\mu(\Phi_0)$ is presented in Fig.~\ref{Fig:C_R_10_50} for a variety of ionic bulk concentrations. There is little qualitative difference in $C_\mu(\Phi_0)$ for the different ion sizes. However, quantitatively, one finds that the smaller the ions, the larger $C_\mu(\Phi_0)$ at a given bulk packing fraction $\eta_b$. It is important to note that ionic bulk concentration $\rho_b$ and packing fraction $\eta_b$, although linearly related, are not the same quantity; small ions have a higher molarity at the same packing fraction and the balance between packing and electrostatics is quite different. Experimentally, one often characterizes bulk concentrations in molarity rather than packing fractions, which renders the ion diameter an important fit parameter. Fig.~\ref{Fig:C_R_10_50} shows that the ion size  can determine the shape of $C_\mu(\Phi_0)$ at a given molarity $\rho_b$. We note, for instance, that the camel-bell crossover packing fraction $\eta_c$ only changes slightly from $\eta_c\approx0.1$ for $d=1$ nm to $\eta_c\approx 0.15$ for $d=0.2$ nm, while the crossover bulk concentration changes by two orders of magnitude from $\rho_c\approx 32$ M for $d=0.2$ nm to $\rho_c=0.16$ M for $d=1$ nm. 
 Fig.~\ref{Fig:C_R_10_50} also shows a bell-shaped capacitance curve for sufficiently concentrated electrolytes, and a camel-to-bell crossover for dilute electrolytes at a sufficiently high electrode potential, with a higher $\Phi^*$ for lower $\eta_b$ and smaller $d$.
The generally lower capacitance for larger ions can be understood qualitatively in terms of the close proximity of smaller ions to the electrode. This is explicit in Eq.~\eqref{Eq:C_phi}, in which $C_\mu$ is inversely proportional to the ion diameter $d$. Hence, a larger ion diameter results in an overall smaller $C_\mu(\Phi_0)$.\\
 
 The results presented in this section evidently show that ion size is a key parameter that largely determines the magnitude of the differential capacitance at a given ionic bulk concentration.
 
\subsection{Second Glance: Asymmetric Electrolytes}
\begin{figure}
     \centering
     \includegraphics[width=\columnwidth]{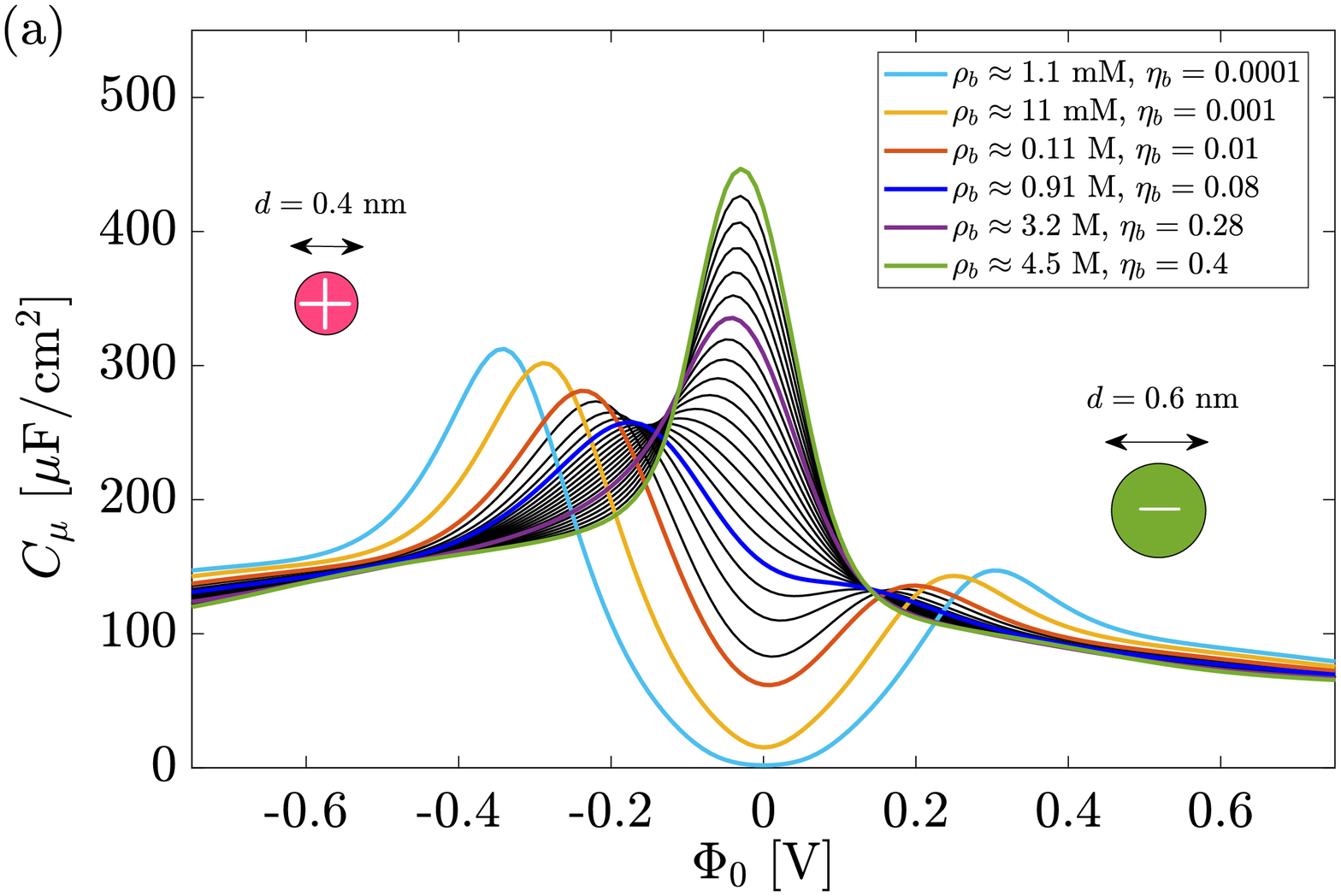}
     \includegraphics[width=\columnwidth]{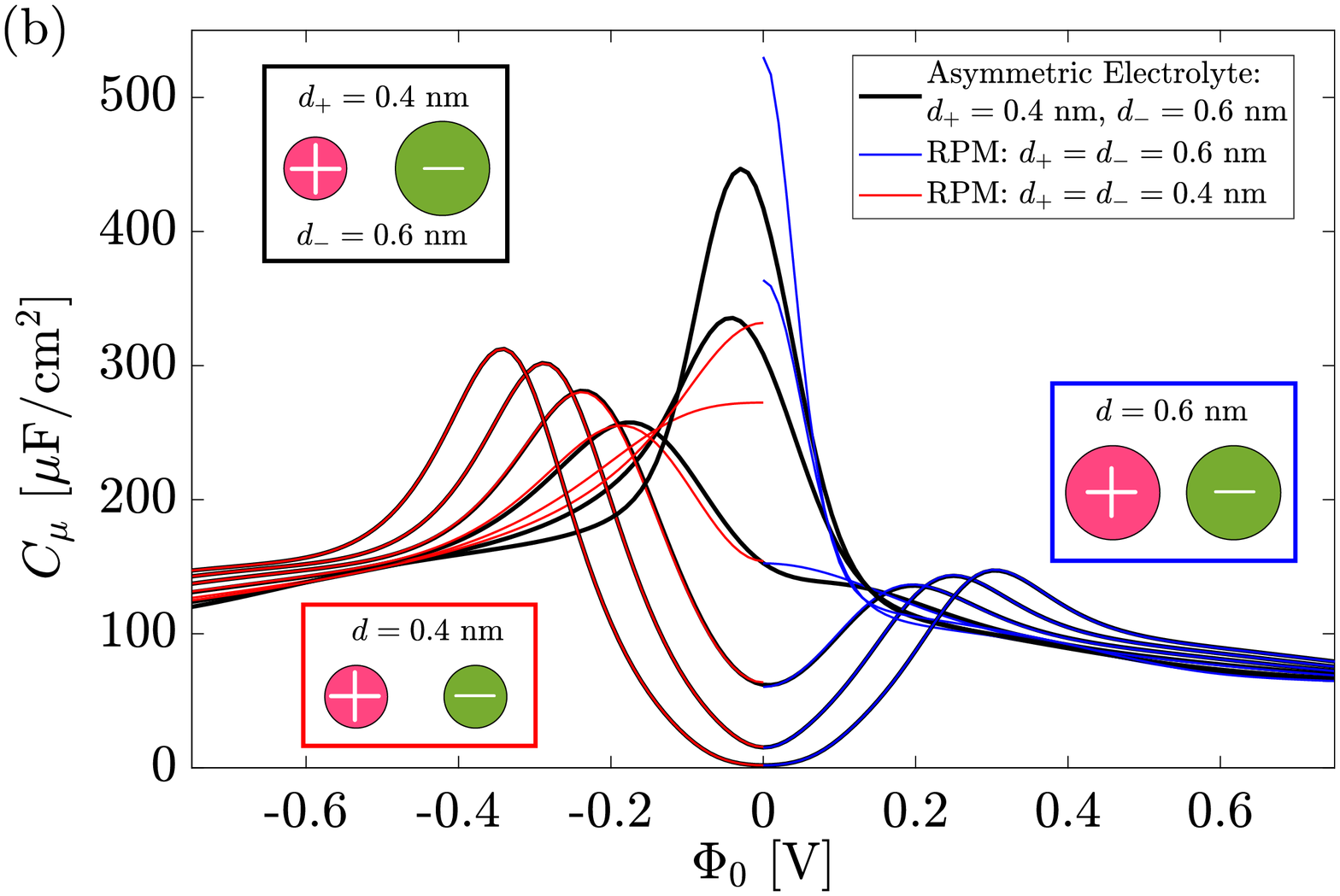}
         \caption{(a) The differential capacitance for an asymmetric electrolyte with $d_+=0.4$ nm and $d_-=0.6$ nm  for a surface separation $H=5$ nm and Bjerrum length $\lambda_B=0.73$ nm. The thick black curves in (b) correspond to the colored ones in (a) and are compared with the differential capacitance of the RPM with $d_+=d_-=0.4$ nm (red) and $d_+=d_-=0.6$ nm (blue) at the same bulk concentration $\rho_b$.}
     \label{Fig:C_R_20_30}
 \end{figure}

Let us now consider an asymmetric electrolyte consisting of cations with diameter $d_+=0.4$ nm and anions with diameter $d_-=0.6$ nm. The voltage-dependent differential capacitance for this system is presented in Fig.~\ref{Fig:C_R_20_30}(a), again for bulk packing fractions ranging from $\eta_b=0.0001$ ($\rho_b\approx1.1$ mM) to $\eta_b=0.4$ ($\rho_b\approx4.5$ M). The asymmetry in the electrolyte is reflected by the asymmetry of Fig.~\ref{Fig:C_R_20_30}(a) w.r.t. $\Phi_0\rightarrow-\Phi_0$; the smaller cations can approach the surface to closer distances than the anions, causing a larger capacitance for negative potentials compared to positive potentials.
The capacitance $C_\mu(\Phi_0)$ at its maximum for the camel-shaped curves at $\Phi_0<0$ is therefore larger than that at $\Phi_0>0$, which motivates the introduction of the notation $\Phi^*_+$ and $\Phi_-^*$ for the maxima located at positive and negative potentials, respectively. For high bulk concentrations with only one maximum (bell-shaped curves), we keep using the notation $\Phi^*$, which is not necessarily close to zero now that the symmetry is broken. In Fig.~\ref{Fig:C_R_20_30} we see that $\Phi^*$ does indeed not vanish in the asymmetric electrolyte, rather it is located at negative potentials, and seems for the present choice of parameters to converge to $\Phi^*\approx -0.03$ V at high bulk concentrations. This also implies that the potential of zero charge takes non-vanishing values. 

In Fig.~\ref{Fig:C_R_20_30}(b) the differential capacitance of the asymmetric electrolyte of Fig.~\ref{Fig:C_R_20_30}(a) is compared to those of the symmetric RPM. We replot the six colored differential capacitance curves of the asymmetric electrolyte of Fig.~\ref{Fig:C_R_20_30}(a) in black now, while in blue ($\Phi_0>0$) and red ($\Phi_0<0$) we plot those for the RPM with $d=0.6$ nm and $d=0.4$ nm, respectively, at the same bulk concentration $\rho_b$. For the three lowest bulk concentrations $\rho_b=1.1$ mM, $\rho_b=11$ mM, and $\rho_b=0.11$ M, we see that the differential capacitance of the asymmetric electrolyte with $d_+=0.4$ nm and $d_-=0.6$ nm (black) is indistinguishable from those of the RPM with $d=0.4$ nm (red) at negative potentials and $d=0.6$ nm (blue) at positive potentials. Hence, $C_\mu$ at negative potentials is fully dictated by the cations, while at positive potentials it is fully dictated by the anions. However, at higher bulk concentrations, close to the camel-bell transition, one finds differences between the RPM-approximation and the actual asymmetric electrolytes, although primarily only for $|\Phi_0|<100$ mV. This can be explained by considering Eq.~\eqref{Eq:Cap_sumrule}, where packing of asymmetric-sized ions in the first layer $\sum_j\rho(d_j/2)$ is evidently different from the RPM. Nevertheless, at higher surface potentials the differential capacitance of the asymmetric electrolyte follows again that of the RPM, because the counterions are fully repelled resulting in a similar packing as in the RPM.

\begin{figure}
     \centering
     \includegraphics[width=\columnwidth]{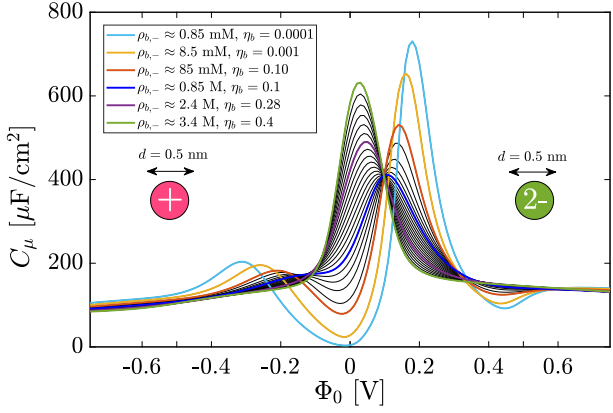}
         \caption{The differential capacitance for an asymmetric 1:2 electrolyte with $d_+=d_-=0.5$ nm. The bulk concentrations for the colored lines are given in the legend and range from $\rho_{b,-}=\rho_{b,+}/2=0.85$ mM to  $\rho_{b,-}=\rho_{b,+}/2=3.4$ M. }
     \label{Fig:C_Z_1_2}
 \end{figure}

\begin{figure}
     \centering
     \includegraphics[width=\columnwidth]{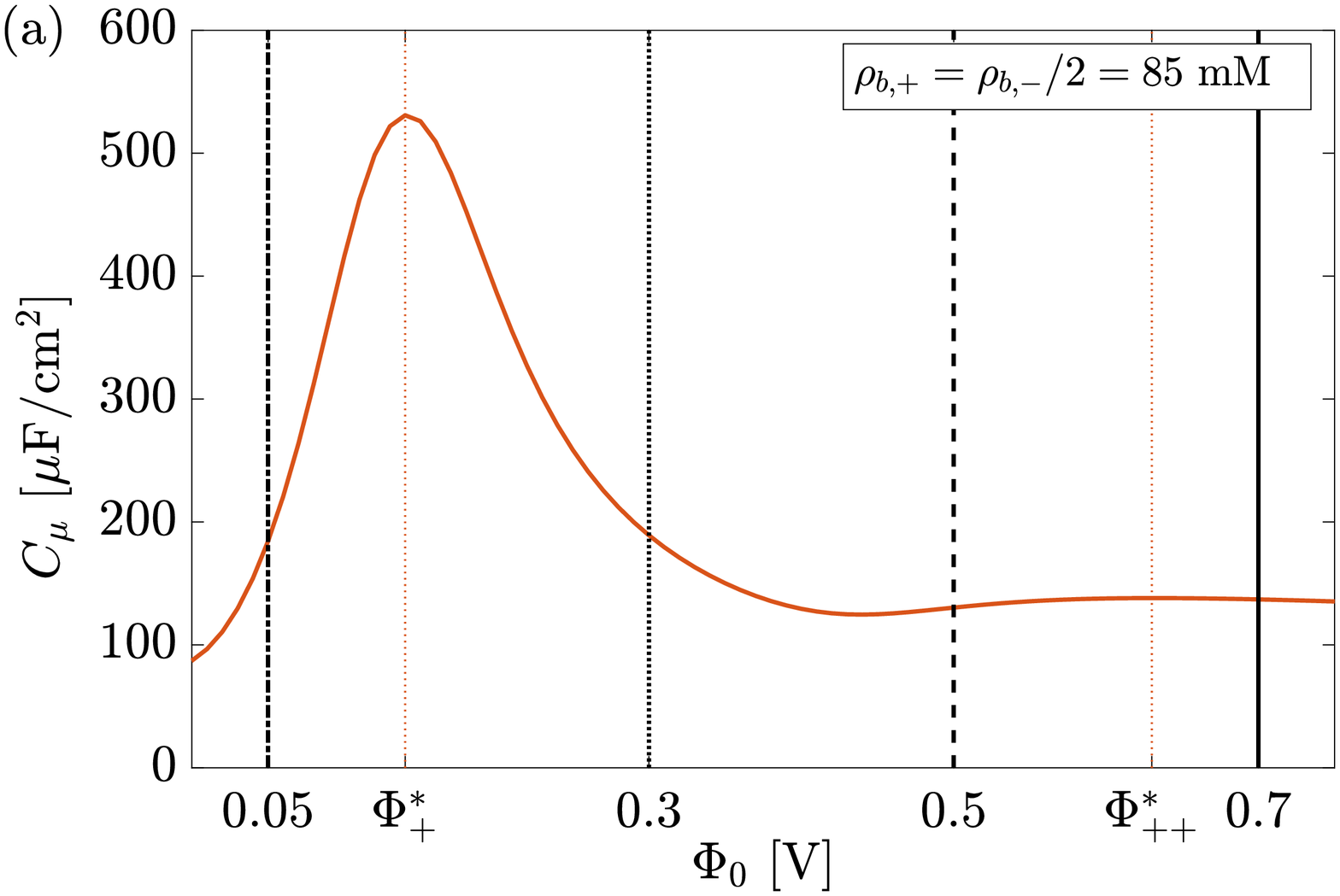}
     \includegraphics[width=\columnwidth]{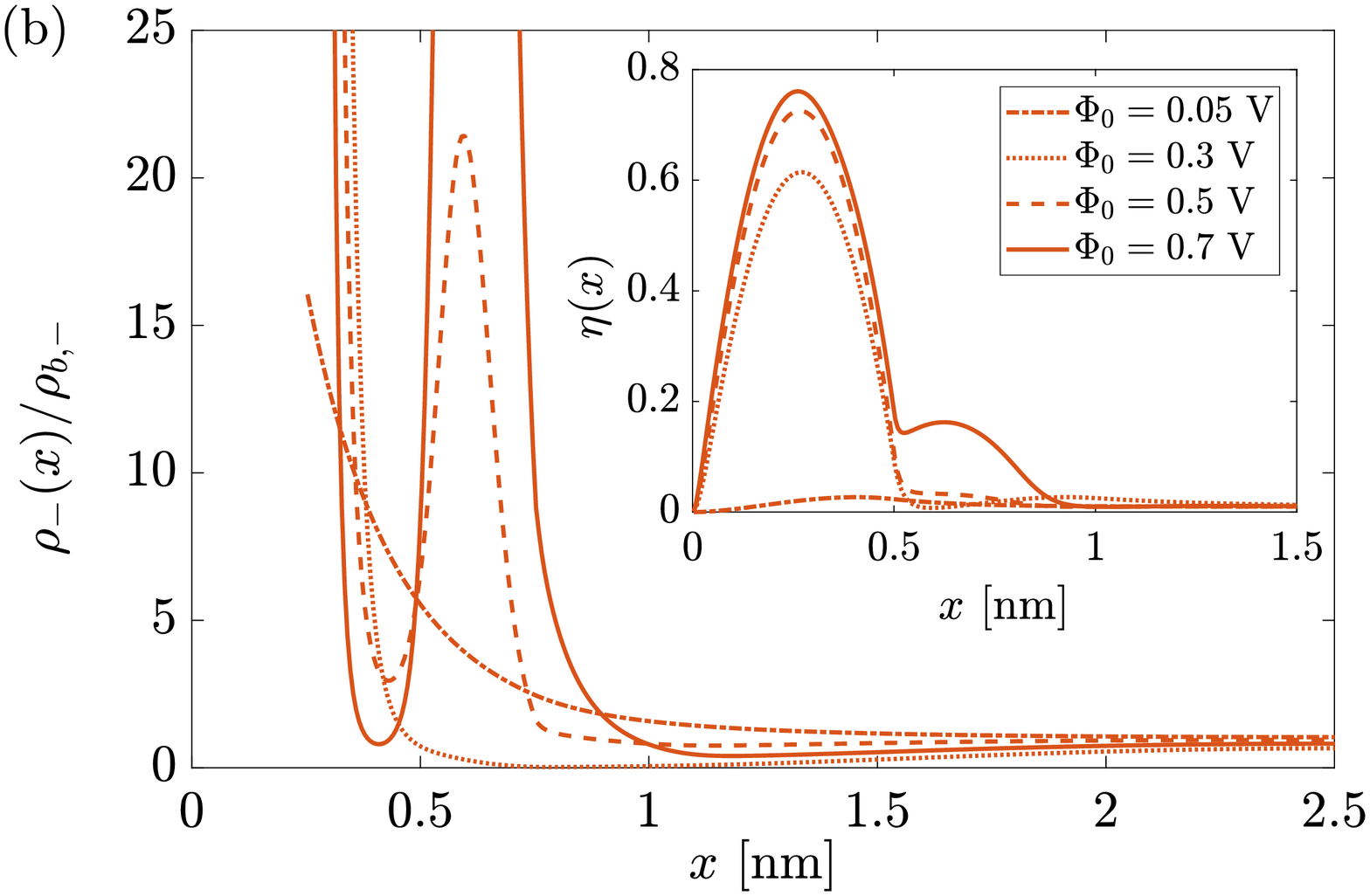}
     \includegraphics[width=\columnwidth]{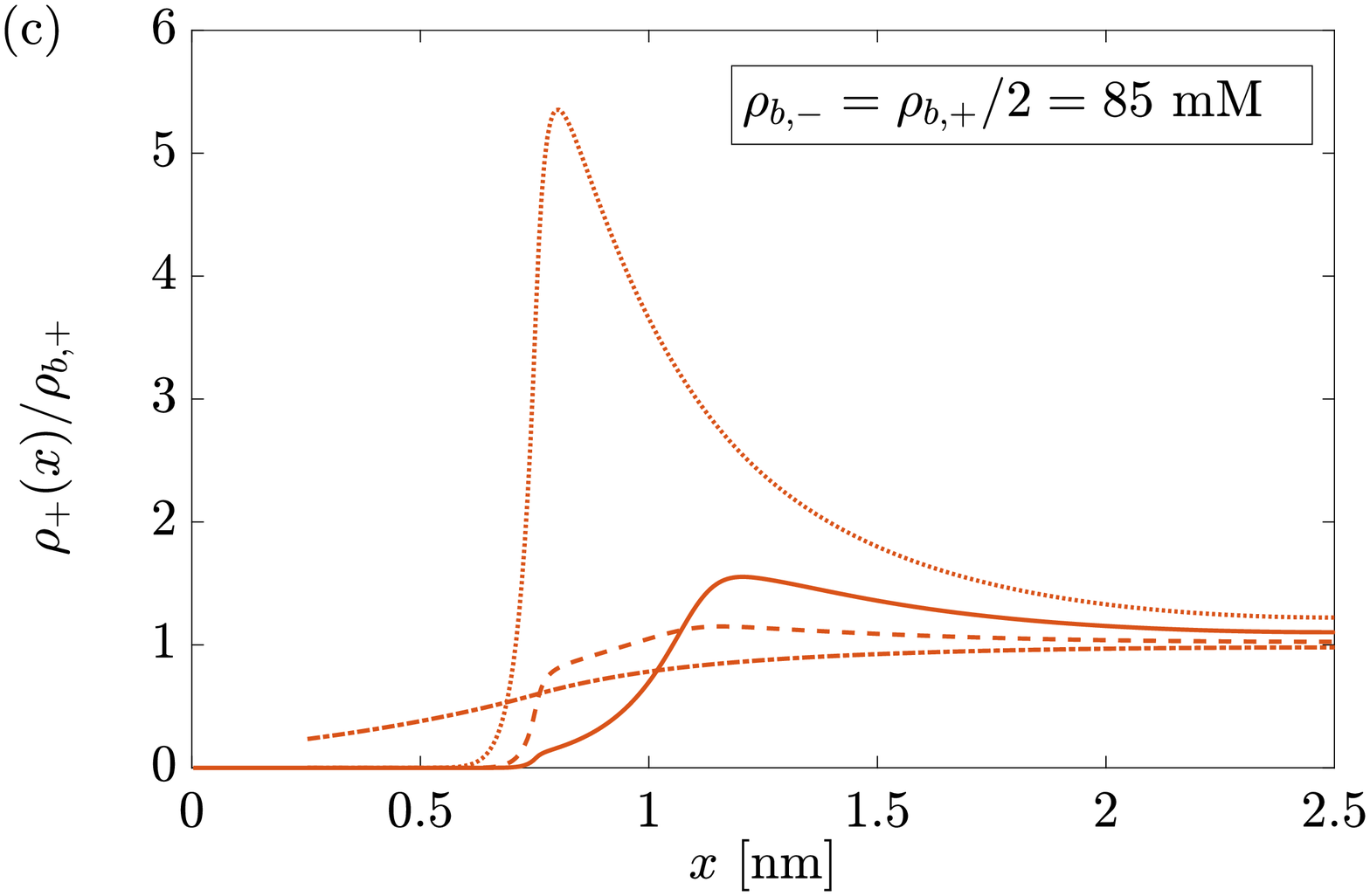}
         \caption{ In (a) the $C_\mu$ curve of the asymmetric 1:2 electrolyte for which the anion and cations density profiles are portrayed in (b) and (c), respectively, at the surface potentials $\Phi_0=0.05$ V (dashed-dotted), $\Phi_0=0.3$ V (dotted), $\Phi_0=0.5$ V (dashed), and $\Phi_0=0.7$ V (solid), as indicated by  the vertical lines in (a). The bulk concentration corresponding to these results is $\rho_{b,+}=\rho_{b,-}/2=85$ mM ($\eta_b=0.01$). The inset in (b) shows the local packing fraction. The $\Phi^*_+$ and $\Phi^*_{++}$ are the surface potentials at which $C_\mu$ has a maximum.}
     \label{Fig:rho_12_asymmetric}
 \end{figure}

Another type of asymmetric electrolyte that we consider is a 1:2 electrolyte in which $z_+=1$ and $z_-=-2$, for convenience with equal diameter $d_+=d_-=0.5$ nm. The capacitance curves for this system are given in Fig.~\ref{Fig:C_Z_1_2}, again for $H=5$ nm and $\lambda_B=0.73$ nm. First let us note that the maximum at positive potentials for bulk concentrations below $\rho_{b,-}=0.85$ M is larger than that of the monovalent RPM at the same packing fraction (see Fig.~\ref{Fig:cap_RPM} or consider the negative potentials). This is a consequence of the higher valency of the anions; fewer ions are needed to screen the charge on the surface and therefore layering/packing requires higher surface potentials. One peculiar feature that was not present in all previous monovalent cases is the minimum capacitance around $\Phi_0\approx0.4$ V (and therefore also a second maximum at $\Phi_0=\Phi_{++}^*\approx 0.6$), which is caused by overscreening~\cite{Kornyshev_2007}.  
At such relative high surface potentials, the positively charged surface attracts the divalent anions to such an extent that it creates a layer of anions with negative charge that exceeds the magnitude of the charge on the surface. Therefore the ions further away from the electrode perceive the  surface as being negatively charged, rather than positive, which causes a net attraction of positive ions. This is shown in Fig.~\ref{Fig:rho_12_asymmetric}(b) and~(c), which presents the anion and cation density profiles, respectively, at the surface potentials indicated by the vertical lines in the capacitance curve of Fig.~\ref{Fig:rho_12_asymmetric}(a), for the bulk concentration of $\rho_{b,-}=\rho_{b,+}/2\approx 85$ mM ($\eta_b=0.1$).  The inset in Fig.~\ref{Fig:rho_12_asymmetric}(b) shows the local packing fraction as defined by Eq.~\eqref{Eq:eta_local}. For low surface potentials $\Phi_0<\Phi^*_+\approx 0.14$ V (dashed-dotted lines) the behaviour is similar to the RPM, but upon increasing the surface potential to $\Phi^*_+<\Phi_0=0.3 $ V$<\Phi^*_{++}$ (dotted line) a clear peak in the cation density profile is formed, as if the surface is negatively charged. Note that in this case, there is still only one dense layer of anions near the surface. Increasing the surface potential to $\Phi_0=0.5$ V (dashed lines) causes a dense second layer of anions, while the cations are repelled. Overscreening is found again at $\Phi_0=0.7$ V $>\Phi^*_{++}$ (solid lines). Hence, the structural changes of the EDL are manifested in the differential capacitance; from a diffuse EDL to overscreening with one layer of cations and anions (up to the first maximum in $C_\mu$), to two layers of anions and a diffuse cation layer (between the first maximum and the first minimum in $C_\mu$), to overscreening with two layers of anions (beyond the second maximum in $C_\mu$). This is also visible in the local packing fraction (see inset in Fig.~\ref{Fig:rho_12_asymmetric}(b)), which shows that a densely-packed first layer is formed for $\Phi_0>\Phi^*_+$ and a second dense layer for $\Phi_0>\Phi^*_{++}$. \\

\subsection{Impurities}
It is extremely difficult to experimentally study an aqueous electrolyte in which only one type of salt is dissolved. Generally, there are always ``impurities", with a much lower bulk concentration than that of the dominant species. Common cations in water are Calcium (Ca$^{2+}$), Magnesium (Mg$^{2+}$), Sodium (Na$^+$), and Potassium  (K$^+$), while the most common anions are Carbonate (CO$_3^{2-}$), Chloride (Cl$^-$), and Sulfate (SO$_4^{2-}$). Although there are techniques to obtain purified (dionized, demineralized) water, extracting all ion types is challenging~\cite{Bassyouni_2019,Gama_2021,Yang_2021}. One particular example would be the presence of divalent ions such as Calcium (Ca$^{2+}$) and Carbonate (CO$_3^{2-}$), which typically have concentrations ranging from below $0.01$ mM (purified water) to $0.5$ mM (recommended amount of Calcium in tap water) to $2$ mM (very hard water)~\cite{WHO,Kozisek_2020}. 

 \begin{figure}
     \centering
     \includegraphics[width=\columnwidth]{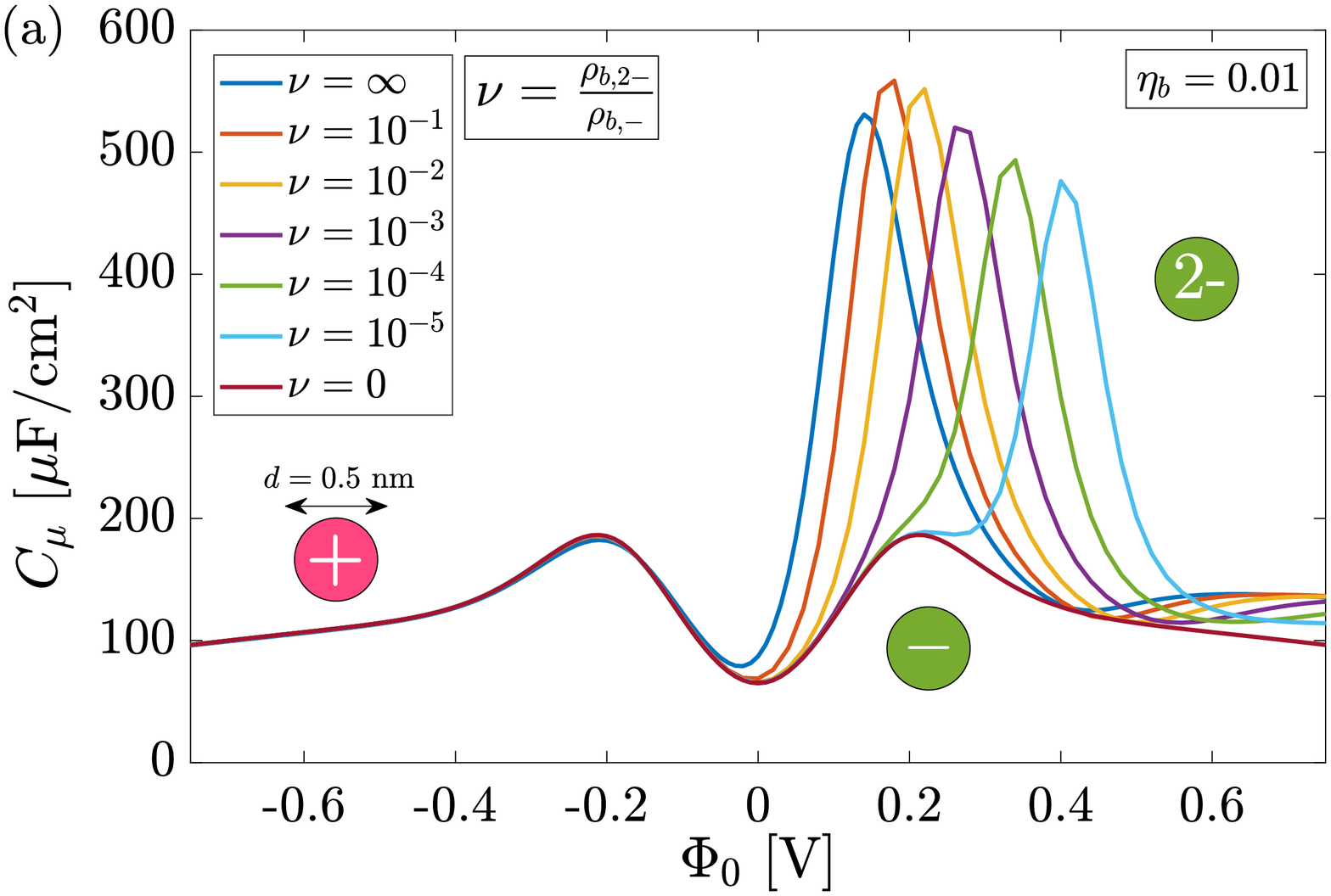}
     \includegraphics[width=\columnwidth]{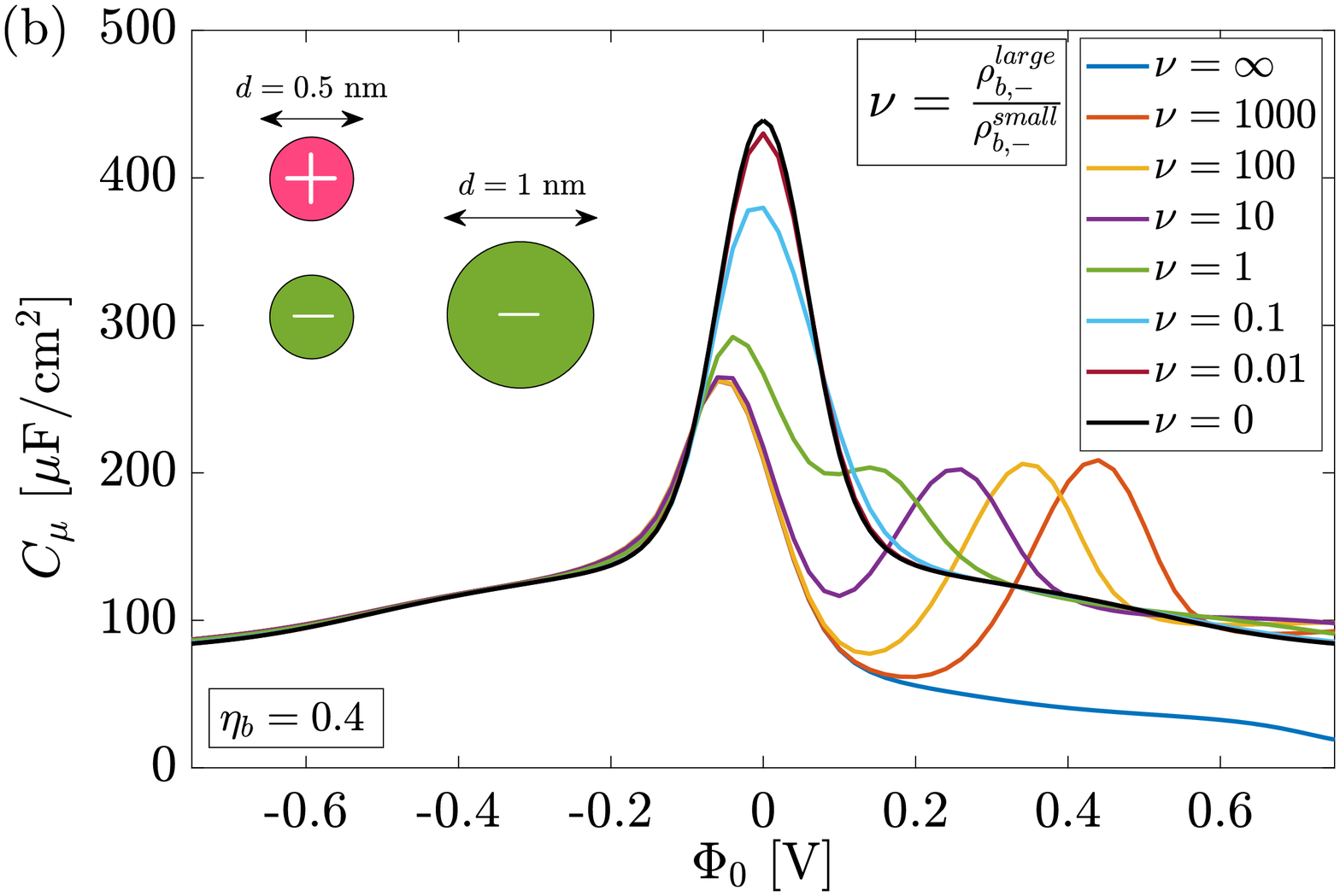}
         \caption{ (a) The differential capacitance for a mixture of anions, in which one species is monovalent and the other divalent. The parameter $\nu$ gives the ratio between the bulk density of those two anion species and ranges from $\infty$ (one divalent anion species) to 0 (on monovalent anion species), while the packing fraction for all curves is fixed at $\eta_b=0.01$. (b) Similar as in (a), but there are two monovalent anions with different size; one being $d_-=d_+=0.5$ nm, and the other $d^{\mathrm{large}}=2d=1$ nm. The parameter $\nu$ is the ration between the bulk density of the large anion w.r.t. small anion, i.e. $\nu=0$ represents the RPM, while $\nu=\infty$ represents the system with one large anion species, while the bulk packing fraction for all curves is fixed at $\eta_b=0.4$.}
     \label{Fig:C_impurity}
 \end{figure}

Let us therefore now consider three-component electrolytes consisting of one cation species and two anion species, with an asymmetry between the anions. First, we consider a 1:1:2 electrolyte of equisized ions with diameter $d=0.5$ nm and $\rho_{b,+}=\rho_{b,-}+2\rho_{b,2-}$ where the anion composition is characterized by $\nu=\rho_{b,2-}/\rho_{b,-}$. The differential capacitance  for this system with a bulk packing fraction $\eta_b=0.1$ is presented in Fig.~\ref{Fig:C_impurity}, where $\nu$ takes the values $\infty$, $10^{-1}$, $10^{-2}$, $10^{-3}$, $10^{-4}$, $10^{-5}$, and $0$, where $\nu=0$ corresponds to a 1:1 electrolyte and $\nu=\infty$ to a 1:2 electrolyte.
Interestingly, even for $\nu=10^{-6}$ (corresponding to $\rho_{b,2-}\approx 1.3\cdot 10^{-6}$ M) we find the divalent anion species to be predominant for surface potentials $\Phi_0>0.3$ V. Again, this can be explained by using Eq.~\eqref{Eq:Cap_sumrule}, which for a 1:2 electrolyte can be written in terms of the effective composition $\tilde{\nu}\equiv\nu\mathrm{e}^{e\beta\Phi(d/2)}$ that indicates whether or not the divalent ions dominate the surface composition, i.e. for $\tilde{\nu}\ll 1$ the monovalent anions dominate while for $\tilde{\nu}\gg 1$ the divalent anions dominate. Because $(e\beta)^{-1}\approx 25$ mV, one finds for  $\Phi_0=0.3$ V $\approx 12/e\beta$ that Boltzmann factor $\exp{(e\beta \Phi(d/2))}$ is large, such that $\mathcal{O}(\tilde{\nu})\approx 1$, even if $\nu=10^{-6}$. That is, the two anion species compete with each other on who may fill the first layer near the surface, and since a divalent ion has a much stronger interaction with the surface charge it wins over the monovalent ion for large surface charges, even if their bulk concentration is much lower. To conclude, even a trace amount of divalent ions, like Ca$^{2+}$ or CO$_3^{2-}$, in a predominant monovalent electrolyte can play a dominant role in the differential capacitance at high surface potentials. 

For two species of anions which differ in size rather than valency, the story is quite different.  Let us again consider a three-component system with monovalent ion species with different ion diameters specified by $d\equiv d_+=0.5$ nm, $d_-^{\mathrm{small}}=d$, and $d_-^{\mathrm{large}}=2d$. Fig.~\ref{Fig:C_impurity} shows the differential capacitance for this system for different anion composition ratios $\nu=\rho_{b,-}^{\mathrm{large}}/\rho_{b,-}^{\mathrm{small}}$ ranging from $\nu=0$ (one anion species with $d_-=d$) to $\nu=\infty$ (one anion species with $d_-=2d$) at a bulk packing fraction of $\eta_b=0.4$. Similar to the previous three-component case with different valencies there is a competition between the two anion species; smaller anions can approach the surface to closer distances than the larger anions, obviously, and therefore the two anion species again compete on who may fill the first layer. For $\nu<1$, the smaller anions dominate at all surface potentials, and Fig.~\ref{Fig:C_impurity}(b) only shows one discernible peak in the capacitance profile. When $\nu$ is increased to $\nu>1$ (the bulk concentration of the larger anions exceeds that of the smaller anions) the larger anions fill the first layer up to a certain surface potential beyond which the smaller anions take over. This point is indicated by a peak in the differential capacitance, after which the smaller anions, although having a much lower bulk concentration, are the dominating factor for the differential capacitance. Again, this can be explained from Eq.~\eqref{Eq:Cap_sumrule}, where it becomes apparent that the competition is between $\rho_-^{\mathrm{small}}(d/2)$ and $\rho_-^{\mathrm{large}}(d)$. For $\nu>1$ and for surface potentials at which the lager anions make up the first layer, the quantity $\rho_-^{\mathrm{large}}(d)$ is similar to the case of which there is only one large cation species. However, when the surface potential is increased such that the smaller ions make up the first layer, then $\rho_-^{\mathrm{large}}(d)$ basically vanishes, due to the presence of the smaller anions that repel the larger ones. Hence, the differential capacitance of the impurity system then follows the curve of the differential capacitance of the system for which there is only one anion species with $d_-=d$. Looking at Fig.~\ref{Fig:C_impurity}, it becomes apparent that a camel-shaped curve can occur at very large packing fractions, although its minimum does not coincide with the potential of zero charge.

\section{Conclusion}
We have carefully investigated the differential capacitance for primitive model electrolytes within classical DFT. A new equation~\eqref{Eq:Cap_sumrule} to interpret the differential capacitance was introduced, which greatly helped to understand its behaviour, because it explicitly showed the importance of the first layer of ions near the surface.  Specifically, it helped to further rationalize the camel-bell crossover, that has been a topic of interest across many studies, as a (smooth) structural change in the density profiles near the electrode. Analogous to peaks in the heat capacity indicating (smooth) changes of the thermal occupancy of microstates, peaks in the differential capacitance indicate structural change. This can either be the forming of dense-packed layers of counterions near the surface, but can also be the reorganisation of layers of cations and anions as we showed for the 1:2 electrolyte with divalent anions, in which overscreening causes a rich structural behaviour that seeped through in the capacitance curves. When the impurities in three-component electrolytes were studied, we again found that the competition for the first layer of ions near the surface determined the behaviour of the capacitance curves. This three-component electrolyte that was studied, with one species of cations and two species of anions in which the relative concentration of the anions was changed, gave much richer physics than perhaps anticipated. Both an asymmetry in the anion valency as well as in the anion diameter was considered. Surprisingly at first, even when the composition of the bulk mixture is very asymmetric, both anion species still have their regime in which they fill up the first layer. For anion asymmetry in the valency, where the relative concentration of the divalent anions is very small, we found that the divalent anions can still fill up the first layer when the surface potential is large enough, due to the much stronger interaction with the charged surface. For diameter asymmetry of the anions, with a low relative concentration of the small anions, we found that they can still fill up the first layer at large surface potential because they can screen the charge on the surface more efficiently. For both cases it is the competition between the two anion species that shapes the differential capacitance. Hence, impurities, as we call them, have a rather strong effect on the differential capacitance. The differential capacitance can thus be used as a probe for the electrolyte composition. Knowing that the differential capacitance is largely determined by the first layer of adsorbed ions, and that peaks in the differential capacitance indicate structural changes, can therefore help to distinguish the components in the electrolyte.

Although Eq.~\eqref{Eq:Cap_sumrule} for the differential capacitance that we derived strictly only holds for the primitive model electrolytes in contact with charged hard walls, the physics that the first layer of ions near the surface dominate its behaviour still holds when the conditions are loosened up. Important is that the ion-ion and ion-surface interaction are strongly repulsive at short separations; our results will be modified if dispersion forces are relevant.

The next step would be to confront the knowledge gained from this analysis to experimental data, allowing interpretation of the measured differential capacitance in terms of EDL structure and bulk electrolyte composition. This could, in turn, lead to a better understanding of the differential capacitance, and consequently to an improved performance of real EDL capacitors. An ultimate goal is to provide knowledge to create sustainable alternatives for Lithium-ion batteries~\cite{Lithium_environment,Lithium_mining}, which would contribute directly to the transition towards renewable energy.

\begin{acknowledgments}
We thank Willem Boon for engaging discussions.
This work is part of the D-ITP consortium, a program of the Netherlands Organisation for Scientific Research (NWO) that is funded by the Dutch Ministry of Education, Culture and Science (OCW). It also forms part of the NWO programme ‘Datadriven science for smart and sustainable energy research’,
with project number 16DDS014. 
\end{acknowledgments}

\appendix

\section{Mean Spherical Approximation} \label{App:MSA}

The electrostatic Helmholtz free energy functional is given by
\begin{align}\label{Eq:F_ex_ES}
   \F_{ex}^{ES}[\{\rho\}]=&\F_{ex}^{MF}[\{\rho\}]+\F_{ex}^{MSAc}[\{\rho\}];\\
     \F_{ex}^{MF}[\{\rho\}] =&\frac{1}{2}\int \dd \rb q(\rb)\Phi(\rb);\\
     \F_{ex}^{MSAc}[\{\rho\}]=&-\frac{k_BT}{2}\sum_{ij}\int\dd \rb\int\dd \rb' \rho_i(\rb)\times\nonumber\\
     &\quad \quad \quad \Delta c^{MSAc}_{ij}(|\rb-\rb'|;\{\rho_b\})\rho_j(\rb'),
\end{align}
where $\Delta c^{MSAc}(\rb,\{\rho_b\})$ is a correction on top of mean-field theory based on the mean spherical approximation (MSA). The potential profile $\Phi(\rb)$ is obtained from solving the Poisson equation
\begin{align}\label{Eq:Poisson}
    \nabla^2 \Phi(\rb)=-\frac{q_{tot}(\rb)}{\eps},
\end{align}
where $q_{tot}(\rb)=q(\rb)+q_{f}(\rb)$ is the total charge density counting both the ionic charge density $q(\rb)=\sum_jz_j\rho_j(\rb)$ and the fixed charges $q_{f}(\rb)$. Solving Eq.~\eqref{Eq:Poisson} for fixed surface potentials in a planar geometry requires the boundary conditions $\Phi(0)=\Phi(H)=\Phi_0$, which give rise to the surface charge densities via $\partial_x \Phi(0+)=-\partial_x \Phi(H^-)=-4\pi\lambda_B\sigma$. Therefore, the electrostatic part of the external potential arises from the Poisson equation and is taken care of within $ \F_{ex}^{MF}[\{\rho\}]$. In the bulk, $\Delta c^{MSAc}(r)$ is given by~\cite{Hiroike}
\begin{align}\label{Eq:app_cMSAc}
    \Delta c_{ij}^{MSAc}(r,\{\rho_b\})=
    \begin{cases}
     c_{ij}^{MSAsh}(r,\{\rho_b\})+z_iz_j\frac{\lambda_B}{r}, \quad  &0\leq r \leq \Delta d_{ij};\\
     c_{ij}^{MSAl}(r,\{\rho_b\})+z_iz_j\frac{\lambda_B}{r}, \quad  &\Delta d_{ij}<r\leq d_{ij};\\
     0, \quad  & r>d_{ij};
    \end{cases}
\end{align}
where $\Delta d_{ij}=|d_i-d_j|/2$ and $d_{ij}=(d_i+d_j)/2$. For $d_i<d_j$ the first term for short (sh) separations reads
\begin{align}\label{Eq:app_cMSAsh}
    c_{ij}^{MSAsh}(r;\{\rho_b\})=2\lambda_B\left[z_iN_j+d_i\zeta(X_i+\frac{1}{3}d_i^2\zeta)\right],
\end{align}
with 
\begin{align}
    X_j=&\frac{z_j-d_j^2\zeta}{1+\Upsilon d_j};\\
    N_j=&\frac{X_j-z_j}{d_j};\\
    \Upsilon=&\pi\lambda_B\sum_j\rho_{b,j}X_j^2;\\
    \zeta=&\frac{1}{H}\sum_j\frac{\rho_{b,j}d_jz_j}{1+\Upsilon d_j};\\
    H=&\sum_j\frac{\rho_{b,j}d_j^3}{1+\Upsilon d_j}+\frac{2}{\pi}\left(1-\eta_b\right).
\end{align}
The second term in Eq.~\eqref{Eq:app_cMSAc} for the longer-ranged part (l) for $\Delta d_{ij}<r\leq d_{ij}$ is given by
\begin{align}
        c_{ij}^{MSAl}(r;\{\rho_b\})=\frac{\lambda_B}{r}\left[A_{ij}+B_{ij}r+C_{ij}r^2+F_{ij}r^4\right];\label{Eq:app_cMSAl}
    \end{align}
        with
    \begin{align}
        A_{ij}=&-\Delta d_{ij}^2\left[\zeta(X_i+X_j)+\zeta^2 d_{ij}^2-N_iN_j\right],\\
        B_{ij}=&-(X_i-X_j)(N_i-N_j)-(X_i^2+X_j^2)\Upsilon-\nonumber\\
        &2d_{ij}N_iN_j+\frac{1}{3}\zeta^2(d_i^3+d_j^3),\\
        C_{ij}=&-\zeta(X_i+X_j)+N_iN_j-\frac{1}{2}\zeta^2(d_i^2+d_j^2),\\
        F_{ij}=&\frac{1}{3}\zeta^2.
\end{align}
These rather involved expressions become more tractable for the monovalent RPM, where $c_{ij}^{MSAsh}(r;\rho_b)=0$, $\zeta=0$ and $\Delta d_{ij}=0$, such that
\begin{align}
    c_{ij}^{MSAl}(r;\rho_b)=&\frac{\lambda_B}{r}\Big[ \Big( -(X_i-X_j)(N_i-N_j)-2X^2\Upsilon\nonumber\\
   & -2dN_iN_j\Big)r    +N_iN_jr^2\Big],
\end{align}
which can be simplified by introducing $D=d+1/\Upsilon$, so that with
\begin{align}
    X_j&=\frac{z_j}{\Upsilon D},\\
    N_j&=-\frac{z_j}{D},
\end{align}
one finds
\begin{align}
    c_{ij}^{MSAl}(r;\rho_b)=z_iz_j\frac{\lambda_B}{r}\frac{-2Dr+r^2}{D^2}.
\end{align}
For a planar geometry, one integrates out the planar coordinates: $c(x)=\int \dd y \dd z c(\sqrt{x^2+y^2+z^2})$.

\bibliography{bibliography}

\section*{DATA AVAILABILITY}

The data that support the findings of this study are available
from the corresponding authors upon reasonable request.

\end{document}